\documentclass[%
 reprint,
nofootinbib,
 amsmath,amssymb,
 aps,
]{revtex4-1}

\usepackage{graphicx}
\usepackage{dcolumn}
\usepackage{bm}
\usepackage{hyperref}
\usepackage{xcolor}
\usepackage{booktabs}
\newcommand{\be}{\begin{equation}}
\newcommand{\ee}{\end{equation}}
\newcommand{\eos}{equation of state}
\newcommand{\eoss}{equations of state}

\begin{document}

\title{Rapid model comparison of equations of state from gravitational wave observation of binary neutron star coalescences}

\author{Shaon Ghosh}
 \email{ghoshs@montclair.edu}
\affiliation{Montclair State University, 1 Normal Ave, Montclair, NJ 07043}
\affiliation{University of Wisconsin-Milwaukee, Milwaukee, WI 53201, USA}

\author{Xiaoshu Liu}%
\author{Jolien Creighton}%
\author{Ignacio Maga\~na Hernandez}
\affiliation{%
 University of Wisconsin-Milwaukee, Milwaukee, WI 53201, USA}%


\author{Wolfgang Kastaun}
\affiliation{Max Planck Institute for Gravitational Physics
(Albert Einstein Institute), Callinstr. 38, D-30167 Hannover, Germany}
\affiliation{Leibniz Universität Hannover, D-30167 Hannover, Germany}

\author{Geraint Pratten}
\affiliation{%
School of Physics and Astronomy and Institute for Gravitational Wave Astronomy, University of Birmingham, Edgbaston, Birmingham, B15 9TT, United Kingdom
}%

\date{\today}

\begin{abstract}
The discovery of the coalescence of binary neutron star GW170817 was a
watershed moment in the field of gravitational wave astronomy.  Among the rich
variety of information that we were able to uncover from this discovery was the
first non-electromagnetic measurement of the neutron star radius, and the cold
nuclear equation of state. It also led to a large \eos{} model selection study
from gravitational-wave data. In those studies Bayesian nested sampling runs
were conducted for each candidate \eos{} model to compute their evidence in the
gravitational-wave data. Such studies, though invaluable, are computationally 
expensive and require repeated, redundant, computation for any new models. 
We present a novel technique to conduct model selection of \eos{} in an 
extremely rapid fashion ($\sim$\,minutes) on any arbitrary model. We test this 
technique against the results of a nested-sampling model selection technique 
published earlier by the LIGO/Virgo collaboration, and show that the results are 
in good agreement with a median fractional error in Bayes factor of about 10\%, 
where we assume that the true Bayes factor is calculated in the aforementioned 
nested sampling runs. We found that the highest fractional error occurs for \eos{} 
models that have very little support in the posterior distribution, thus resulting in 
large statistical uncertainty. We then used this method to combine multiple binary 
neutron star mergers to compute a joint-Bayes factor between \eos{} models. 
This is achieved by stacking the evidence of the individual events and computing 
the Bayes factor from these stacked evidences for each pairs of \eos. 
\end{abstract}

\pacs{Valid PACS appear here}
\maketitle


\section{Introduction}
\label{sec:intro}
Neutron stars (NS) are the densest objects that are known to exist in the
universe. The structure of isolated nonrotating NS is completely
determined by the Tolman-Oppenheimer-Volkoff (TOV) equations  \cite{PhysRev.55.374}. 
To solve the TOV equations, however, one needs to know the barotropic {\em
equation of state} which connects the pressure $(p)$ of nuclear
matter at energy density $(\epsilon)$:
\be
\label{eq:EquationOfState}
p = p(\epsilon)\,.
\ee
The \eos{} of cold matter at extreme density is expected to be universal, i.e.,
all NS are expected to share the same \eos. Thus, the equation of
state is a fundamental relation of great importance for understanding NS, and
more generally, matter at extreme densities. In a typical NS, the central
density can, however, reach values as high as ten times the nuclear saturation
density \cite{annurev-astro-081915-023322}.  Such extreme environments cannot
be emulated in laboratories. Moreover, limitations in our knowledge of the
strong nuclear force, which is partially responsible for the pressure resisting
the collapse of a NS due to its gravity\footnote{the degeneracy pressure being
the other component responsible for the hydrostatic equilibrium of the star.},
results in inadequate theoretical models of matter at such densities. Thus, the
lack of experimental data at high densities and incomplete  theoretical
understanding of NS models both contribute to the uncertainty in the modeling
of the nuclear equation of state. These uncertainties have led to the
development of numerous of NS \eos{} models. 

Solving the TOV equation for these \eos{} models allows us to compute the radius
of stable nonrotating NS as function of mass. Moreover,  it can be
shown that there exists one-to-one map between the $p(\epsilon)$ relations of
a NS and its mass-radius relations, i.e., there is a one-parameter family of
NS based on central pressure \cite{1992ApJ...398..569L}.
Observations of pulsars and measurements of their mass and radius provide us
with valuable information about the NS \eos. Traditionally, this had
been the primary avenue of investigation in this field.  However, the vast
majority of known galactic pulsars ($\sim 3000$) are isolated and only around
10\% of systems exist in binaries \cite{Manchester_2005, NSCat}, for which
precise mass measurements are possible.  The measurement of a NS's radius 
using electromagnetic observation is quite challenging. Most radius measurement 
techniques for NSs are either based on detecting surface thermal emission, 
using spectroscopic data to infer the effective temperature assuming isotropic 
emission models. Alternatively, this could involve the measurement of the general 
relativistic effects of the NS's gravity on the thermal emission, for 
example by projects like Neutron Star Interior Composition Explorer (NICER)
\cite{NICER, Miller:2019cac, Riley_2019}. However, all these techniques are
model-dependent and have multiple sources of systematics
\cite{annurev-astro-081915-023322}.

Detection of gravitational waves from binary neutron stars (BNS) using ground-based
gravitational wave detectors like LIGO \cite{TheLIGOScientific:2014jea} and 
Virgo \cite{TheVirgo:2014hva} presents us with a completely new way of 
measuring the NS equation of state. It allows us to measure the masses of the 
NSs directly, and electromagnetic emission model independent measurement of the star's 
structure via the tidal deformability. In a coalescing BNS system the spin \cite{PhysRevD.57.5287} and 
tidal interactions between the two stars will lead to a change in their shape, and 
hence changes the quadrupole moment of the binary system. The time-changing 
induced quadrupole moment results in a faster inspiral as more orbital energy 
goes into radiation and stellar distortion and this results in observable effects 
in the gravitational waveform. If the star is subject to a quadrupolar field given by
$\mathcal{E}_{\rm ij}$ then the resulting quadrupole moment due to tidal
interactions is given by 
\be
\label{eq:tidal-quad-field}
Q_{ij} = -\lambda \mathcal{E}_{ij}\,,
\ee
where $\lambda$ is called the tidal deformability of the NS, which is related
to the \eos{} dependent tidal Love number $k_2$ \cite{Hinderer_2008} as follows
\be
\label{eq:tidal-love-number}
\lambda = \frac{2}{3G}k_2 R^5
\ee
where $R$ is the radius of the NS (which, for a given mass, also
depends on the \eos).
For a given \eos{} and mass of the NS, the tidal deformability,
$\lambda$ can be computed by solving the TOV equations along with a differential equation 
obtained from combining metric perturbation due to external quadrupolar tidal field
\cite{Flanagan:2007ix, Hinderer_2008}. Therefore, any measurement of the tidal
deformability will also lead to constraints on the NS equation of state. This
is where gravitational-wave data for coalescing binaries of NSs provide 
us with an excellent observational window for nuclear matter measurements. 
By modeling the gravitational waveform while taking into account
the tidal deformabilities as extra model parameters \cite{Read:2009yp, DelPozzo:2013ala,
Agathos:2015uaa}, it is possible to obtain the posterior distribution of the NS masses and tidal 
deformabilities from Bayesian parameter estimation 
\cite{van_der_Sluys_2009, PhysRevD.81.062003, Raymond_2010, Rodriguez_2014,
PhysRevD.91.042003} to get posterior distribution on the NS masses and tidal
deformabilities.  This was conducted for the case of GW170817 by the LIGO/Virgo Collaboration
\cite{PhysRevX.9.011001}. Additionally, the constraints on the radii of the NSs were also
estimated \cite{PhysRevLett.121.161101, PhysRevLett.121.091102}
after imposing an \eos{} insensitive relation between the mass, radius, and tidal deformability.
Furthermore, it is possible to parametrize the \eos{} using piecewise polytropes
or a spectral representation \cite{PhysRevD.79.124032, PhysRevD.82.103011} to
infer the \eos{} parameters for a given representation
\cite{PhysRevLett.121.161101}. Finally, nonparametric inference of the cold
NS \eos{} \cite{Landry:2018prl, Landry:2020vaw} can help in relaxing
the choice of parametrization in describing the \eos.

All of these aforementioned studies, which are either \eos{} agnostic, or which
are attempting to reveal the NS \eos{} with minimal (albeit varying degrees of)
assumptions, are extremely important. One can, however, also make a good
argument of studying various \eos{} models in the literature that are based on 
nuclear theory. Analysis of these \eos{} models can give us insight into 
these various theories. There are a few studies in constraining of \eos{} models that 
are motivated from nuclear physics theory \cite{Capano:2019eae, Dietrich1450}, and a 
large-scale \eos{} model selection study has also been conducted by the LIGO/Virgo 
Collaboration \cite{LIGOScientific:2019eut} which has investigated 24 \eos~models
using Bayesian model selection techniques. Bayesian model selection of NS \eos{} requires 
conducting Bayesian parameter estimation on gravitational-wave data using each model 
as a prior. Conducting such studies for multiple \eos{} is computationally expensive. 
There is much redundancy in such studies which we want to avoid so that any new \eos{} model 
can be rapidly tested against any other (new or existing) models. We present in this paper a 
novel technique that allows for reliable and rapid computation of Bayes factors between 
\eos{} models estimated from a single \eos{} agnostic Bayesian parameter estimation run 
on the gravitational-wave data.

The paper is organized as follows: in Sec.~\ref{sec:eos_model_selection}, we
briefly review the technique of Bayesian model selection for NS \eos~using 
gravitational-wave data. In Sec.~\ref{sec:approxevidencederivation} we introduce 
the approximate method for Bayes factor estimation that we use to perform rapid 
model selection. In Sec.~\ref{sec:pe-gw190425} we give the details of the 
parameter estimation runs that are necessary for the application of the 
approximation method. Then in Sec.~\ref{sec:results} we
discuss the results of this technique, where we first show how the method
performs compared to the full Bayesian analysis in
\cite{LIGOScientific:2019eut}, and then in \ref{sec:stacking} we use this
technique to demonstrate its ability to stack Bayes factors from multiple
BNS mergers in order to make a joint model selection statement.
Finally, in Sec. \ref{sec:stacking_real} we apply the method of stacking to
obtain a joint Bayes factor between models of NS \eos{} as informed by
the two observed BNS coalescences, GW170817 and GW190425 
\cite{RICHABBOTT2021100658}.

\section{Model selection of equation of state}
\label{sec:eos_model_selection}
The time-varying strain at an interferometer due to gravitational wave from a
coalescing compact binary system is given by $h(t; m_1, m_2, \lambda_1,
\lambda_2, \vec{\theta} )$, where $\vec{\theta}$ is comprised of the spins of 
the binary, and extrinsic parameters of the source such as its sky-position, the angle 
of inclination of the binary, the gravitational wave polarization angle, distance to 
the source, and its phase at coalescence. For this work we have ignored spin 
contributions to the tidal deformabilities and maximum masses of the binary components.
The quantities $m_i$ and $\lambda_i$, where $i=1, 2$, are the masses and 
tidal deformabilities of the two compact objects. The NS \eos~gives us a map between 
the masses and the tidal deformability parameters: $\lambda = f(m)$. The interferometer 
data $d$ in presence of the gravitational wave signal can be expressed as 
$d(t) = h(t) + n(t)$, where $n(t)$ is the noise in the detector. We can define a likelihood 
function in the parameter space defined by $(m_1, m_2, \lambda_1, \lambda_2, \vec{\theta})$ 
for the observed data as $\mathcal{L} = \mathcal{L}(d \mid m_1, m_2, \lambda_1, \lambda_2,
\vec{\theta})$, which can be used to construct the evidence for the equation of state model 
defined by $f$,
\begin{widetext}
\be
\label{eq:evidence}
\mathcal{Z}_f = \int \mathcal{L}(d\mid m_1', m_2', \lambda_1', \lambda_2', \vec{\theta}')\,p(m_1', \lambda_1', m_2', \lambda_2' \mid f)\, p(\vec{\theta}')\, dm_1'\, dm_2'\, d\lambda_1'\, d\lambda_2'\,d\vec{\theta}'\,,
\ee
\end{widetext}
where the prior for the specific \eos{} model is encoded in the probability
$p(m_1, \lambda_1, m_2, \lambda_2 \mid f)$. Computing the value of $\mathcal{Z}_f$ and $\mathcal{Z}_g$ for
two \eos{} models $f$ and $g$ and then taking their ratio will give us the Bayes factor $\mathrm{BF}^f_g=\mathcal{Z}_f/\mathcal{Z}_g$
between the two models. This technique was employed in
\cite{LIGOScientific:2019eut} and was used to investigate Bayes factors between
a multitude of \eoss{} using the information available from the interferometer
data around the time of GW170817 \cite{gw170817}. For each of these \eoss{} the 
evidence in Eq.~(\ref{eq:evidence}) was evaluated by conducting a multidimensional 
Bayesian parameter estimation using nested sampling \cite{2004AIPC..735..395S}.  
This method, however, is computationally expensive. In the study conducted 
in \cite{LIGOScientific:2019eut}, the computation of evidence for each \eos~ took 
$\gtrsim$1\,week  to finish. Furthermore, if one intends to compare any new \eos{} 
against the existing models, a fresh Bayesian analysis will need to be conducted. 
This renders the method impractical in the long run (particularly when it must be 
repeated over a number of gravitational wave events to obtain a joint Bayes factor 
as we describe in Sec.~\ref{sec:stacking}).  Whenever a new \eos{} models needs 
to be included in the study, not only new parameter estimation runs need to be 
conducted for the new events, but for all events in the past. To address this issue 
we have developed a method of an evidence-approximation scheme that allows the 
use of a single Bayesian parameter estimation result to compute the Bayes factor 
between any two arbitrary \eos{} models in a very rapid fashion (typically within 
minutes\footnote{about an hour if uncertainties need to be calculated}).  This method 
allows the user to define any new \eos{} model and compare that with any existing 
model or another new model without having to repeat the full parameter estimation 
runs. Instead of using the aforementioned parametrization of 
$(m_1, m_2, \lambda_1, \lambda_2)$ we re-parametrize the likelihood to 
$(\mathcal{M}, q, \tilde{\Lambda}, \delta\tilde{\Lambda})$, where 
$\mathcal{M} = (m_1m_2)^{3/5}/(m_1 + m_2)^{1/5}$ is the chirp mass of the
binary, $q = m_2/m_1$ is the mass ratio, assuming the convention $m_1>m_2$.
$\tilde{\Lambda}$ and $\delta\tilde{\Lambda}$ are tidal parameters defined as
\begin{widetext}
\begin{eqnarray}
\label{eq:lambda_lambdat_conversion}
 \tilde{\Lambda}  =  \frac{8}{13} \left[ (1 + 7\eta - 31\eta^2)(\Lambda_1 + \Lambda_2) +  \sqrt{1 - 4\eta}\,(1 + 9\eta - 11\eta^2)(\Lambda_1 - \Lambda_2)\right] 
\end{eqnarray}
\begin{eqnarray}
\label{eq:lambda_dlambdat_conversion}
\delta\tilde{\Lambda} & = & \frac{1}{2} \left[\sqrt{1 - 4\eta} \left(1 - \frac{13272}{1319}\eta + \frac{8944}{1319}\eta^2\right)(\Lambda_1 + \Lambda_2) + \left(1 - \frac{15910}{1319}\eta + \frac{32850}{1319}\eta^2 + \frac{3380}{1319}\eta^3\right)(\Lambda_1 - \Lambda_2)\vphantom{\sqrt{1 - 4\eta}}\right]\,,
\end{eqnarray}
\end{widetext}
where, $\eta = m_1m_2/(m_1+m_2)^2$ is the symmetric mass ratio, and 
$\Lambda_1 = G\lambda_1 [c^2 /(Gm_1)]^5$ and $\Lambda_2 = G\lambda_2 [c^2 /(Gm_2)]^5$ 
are dimensionless tidal deformabilities of the two stars,
as described in \cite{Flanagan:2007ix, Favata:2013rwa, Wade:2014vqa}. This
allows us to neatly isolate the dominant tidal contribution to the
gravitational waveform, $\tilde{\Lambda}$, from the subdominant contribution of
$\delta\tilde{\Lambda}$.  We describe the details of the evidence approximation
scheme in the next section.

\section{Evidence approximation scheme}
\label{sec:approxevidencederivation}

Bayesian model selection among a number of \eos{} models requires the
computation of the evidence $\mathcal{Z}$ for each model under consideration.
The evidence can be computed by methods such as nested sampling~
\cite{2004AIPC..735..395S}, but this multidimensional integral must be 
recomputed for each \eos{} model. Here we describe an approximation 
scheme by which the evidence for each model can be obtained from a 
low-dimensional integral over a marginalized likelihood function that is 
constructed from a set of samples drawn from a distribution with a single 
MCMC sampling.

The approximation method exploits the fact that the chirp mass of a
BNS system is extremely well measured from the gravitational
wave signal, whereas a second mass parameter, e.g., the mass ratio $q$,
will be significantly less constrained.  Additionally, the tidal parameter
$\tilde{\Lambda}$, first enters the post-Newtonian expansion of the waveform
at the 5th post-Newtonian order, while $\delta\tilde{\Lambda}$ first enters
the expansion at the 6th post-Newtonian order.  The impact of
$\delta\tilde{\Lambda}$ is much weaker than the one of $\tilde{\Lambda}$ and
can often be neglected.

If we ignore spin contributions, a given \eos, $E$, defines how the tidal
deformabilities are related to the NS mass, $\Lambda_E(m)$, as well as
a maximum mass $m_{\max,E}$ of a nonrotating NS\@.  We assume
that a compact object with $m>m_{\max,E}$ is a black hole (ignoring
the possibility of rapidly rotating supramassive NSs) and define
$\Lambda_E(m)=0$ for $m>m_{\max,E}$.  The function $\Lambda_E$ then
allows us to obtain the functions $\tilde{\Lambda}_E(\mathcal{M},q)$
and $\delta\tilde{\Lambda}_E(\mathcal{M},q)$ using Eqs. \ref{eq:lambda_lambdat_conversion}
and \ref{eq:lambda_dlambdat_conversion}, 
which determine the parameters $\tilde\Lambda$ and $\delta\tilde\Lambda$ for a BNS
with mass parameters $(\mathcal{M},q)$~\cite{Wade:2014vqa}.
The evidence for \eos{} $E$ is then
\begin{widetext}
\begin{equation}
    \mathcal{Z}_E = \int
    \mathcal{L}(d \mid \mathcal{M}', q', \vec{\theta}',
        \tilde\Lambda_E(\mathcal{M}', q'),
        \delta\tilde\Lambda_E(\mathcal{M}', q'))
    \,p(\mathcal{M}',q')\,p(\vec{\theta}')\,d{\mathcal{M}}'\,dq'\,d\vec{\theta'}
\end{equation}
and if we marginalize over the parameters $\vec{\theta}$ and express
the \eos{} constraints as delta-functions, we have
\begin{equation}\label{e:evidence_delta}
    \mathcal{Z}_E = \int
    \mathcal{L}(d \mid \mathcal{M}', q', \tilde\Lambda', \delta\tilde\Lambda')
    \,\delta(\tilde\Lambda'-\tilde\Lambda_E(\mathcal{M}', q'))
    \,\delta(\delta\tilde\Lambda'-\delta\tilde\Lambda_E(\mathcal{M}', q'))
    \,p(\mathcal{M}',q') \,d{\mathcal{M}}'\,dq' \,d\tilde\Lambda' \,
    d\delta\tilde\Lambda'\,.
\end{equation}
\end{widetext}
The prior distribution over masses $p(\mathcal{M},q)$ is taken to be the
same for any \eos{} in our approximation, which would neglect astrophysically
motivated priors involving mass gaps between the most massive NSs and the 
least massive black holes.

To obtain the marginal likelihood
$\mathcal{L}(d \mid \mathcal{M}', q', \tilde\Lambda', \delta\tilde\Lambda')$
we use well-tested and well-maintained parameter estimation code to compute
a posterior distribution:
Application of Bayes's theorem reduces the marginal likelihood to a
posterior distribution
\begin{equation}\label{e:posterior_likelihood}
    p(\mathcal{M},q,\tilde\Lambda,\delta\Lambda \mid d)
    \propto p(\mathcal{M},q)
    \mathcal{L}(d \mid \mathcal{M}, q, \tilde\Lambda, \delta\tilde\Lambda)
\end{equation}
where a uniform prior over $\tilde\Lambda$ and $\delta\tilde\Lambda$ is
imposed.  This prior in the tidal deformabilities is not physically
meaningful: its seemingly unphysical form is needed only to form the correct
relation between the posterior distribution and the marginalized likelihood.
(The prior over the mass parameters, however, is physically relevant.)
However, it must be noted that the prior over deformabilities needs to have
support over all possible deformabilities allowed by the \eos{} under
consideration.
As the equation of state dependence appears only in the delta-functions
in Eq.~(\ref{e:evidence_delta}), the constant of proportionality
in Eq.~(\ref{e:posterior_likelihood}) is not \eos{} dependent and so is
irrelevant for our purposes.

As mentioned earlier it is observed that
the posterior distribution is sharply-peaked about some
well-measured chirp mass $\mathcal{M}_0$, and is largely independent of
the subdominant tidal parameter $\delta\tilde\Lambda$, so, to a very
good approximation
\begin{equation}
    p(\mathcal{M},q,\tilde\Lambda,\delta\tilde\Lambda \mid d)
    \propto p(q, \tilde\Lambda \mid d)\,\delta(\mathcal{M}-\mathcal{M}_0)
\end{equation}
where $p(q, \tilde\Lambda \mid d)$ is the posterior distribution marginalized
over all parameters apart from $q$ and $\tilde\Lambda$.
Performing the integrals in the evidence, we find
\begin{equation}
    \label{e:1dlineint}
    \mathcal{Z}_E \propto \int p(q',\tilde\Lambda_E(\mathcal{M}_0,q') \mid d)
    \,dq'
\end{equation}
i.e., proportional to the line integral over the two-dimensional posterior
distribution $p(q, \tilde\Lambda \mid d)$ along the curve
$\gamma : q \to \tilde\Lambda_E(q,\mathcal{M}_0)$ determined by the
measured chirp mass $\mathcal{M}_0$ and the \eos{} $E$.  Again, the
constant of proportionality is not \eos{} dependent and so is irrelevant
when computing Bayes factors between different \eoss.  The curve
$\gamma$ contains the full support of the prior on $q$ which may include
values where one or both of the components exceed $m_{\max,E}$, which
correspond to neutron-star black-hole binaries and binary black holes.

In our work, we used MCMC stochastic sampling to draw samples from the
posterior distribution.  We then used a kernel density estimator (KDE)
of samples $(q,\tilde\Lambda)$ drawn from the posterior to obtain
a KDE estimate $K(q,\tilde\Lambda)$ of the function $p(q,\tilde\Lambda\mid d)$.
The line integral is then written in terms of this estimate as
\begin{equation}
    \label{e:1dlineintkde}
    \mathcal{Z}_E \propto \int K(q',\tilde\Lambda_E(\mathcal{M}_0,q')) \,dq'\,.
\end{equation}
Within this approximation scheme, samples from the posterior distribution
$p(q,\tilde\Lambda\mid d)$ only needs to obtained once (per event)
using an \eos{} agnostic prior that is uniform over $\tilde\Lambda$
and $\delta\tilde\Lambda$.  Once this has been computed, the evidence
for any \eos{} model can be efficiently obtained by a one-dimensional line
integral of Eq.~(\ref{e:1dlineintkde}).
Because we approximate the posterior distribution with a KDE based on a
finite number of samples, this estimation method is subject to sampling
uncertainty.  We estimate the uncertainty via a bootstrapping approach.
From the kernel density we resample the same number of points as the
original samples and recompute the evidence using this new set of points.
This should give us another instance of the evidence for the same event.
Continuing this multiple times we are able to create a distribution of the
evidence, which in turn will give us an estimate in the uncertainty of
our Bayes factors between two \eos{} models.  While we believe any biases
associated with the point estimate of the evidence ratio to be small, the
bootstrapping procedure introduces a bias due to oversmoothing of the
posterior.  We therefore include a factor of two based on simulation studies
(see Appendix)
in the error estimates shown in Sec.~\ref{sec:results} in order to render
conservative error estimates.

For the results in this work we use
a prior that is uniform in $0\le\tilde\Lambda\le3000$ and
$-500\le\delta\tilde\Lambda\le500$.  Note that this formally includes
(unphysical) negative values of the individual tidal deformabilities
$\Lambda_1$ and $\Lambda_2$.  These negative values can lead to issues
with waveform generators that fail upon encountering these unphysical
points.  In this study we used the TaylorF2 waveform \cite{Bini:2012gu} truncated at 5th
post-Newtonian order which is a post-Newtonian nonprecessing frequency domain 
waveform that includes tidal effects. TaylorF2 does not depend on $\delta\tilde\Lambda$ 
and so was largely immune to this issue.  The \eos{} will affect the gravitational 
waveform at high frequencies near the point at which the NSs come into contact 
that are not modeled by the TaylorF2 waveforms.  In this work we chose to 
terminate the waveform at the innermost stable circular orbit (ISCO) \cite{Wade:2014vqa}. 
As the detectors are not sensitive at such high frequencies, this arbitrary termination 
condition is unimportant. A more general approach would be to use an arbitrary 
prior $p(\Lambda_1,\Lambda_2)$ with support only for $\Lambda_1\ge0$ and 
$\Lambda_2\ge0$ and to divide the posterior distribution by the marginalized 
prior $p(\tilde\Lambda)$ (which can be done by reweighting samples drawn 
from the posterior distribution).

\section{Equation of state agnostic parameter estimation runs}
\label{sec:pe-gw190425}
The evidence approximation post-processes the Bayesian posterior samples. The
current implementation of the method uses an equation of state agnostic MCMC
exploration of the 11 dimensional parameter space spanned by $(\mathcal{M}, q,
\tilde{\Lambda}, \delta\tilde{\Lambda}, \chi_1, \chi_2, d_L, \theta_{JN}, \psi,
t, \delta_c)$. For GW170817 we fix the sky-position at the location of NGC4993.
We use \textsc{\texttt{LALInference\_MCMC}} from LALsuite library
\cite{lalsuite} to obtain posterior samples for $\tilde{\Lambda}$ and $q$ as
well as the other aforementioned parameters. It implements Metropolis–Hastings
algorithm \cite{{Metrolis}, {Hastings}} with parallel tempering
\cite{{PhysRevD.81.062003}, {PhysRevD.91.042003}}, which modifies the
likelihood function $p(d\mid\theta)$ to $p(d\mid\theta)^{1/T}$ for different
`temperature' ($T$) chains. In all the examples that follow we choose to use 8
different temperatures from $T=1$ to $T=50$ for each MCMC run. Parallel tempering
allows, at higher temperatures efficient `global' exploration of the
prior-space, while at the lower temperatures, finer detailed exploration of
local space around regions of higher likelihood. During the sampling, each
chain swaps samples periodically based on the criteria presented in Ref.
\cite{PhysRevD.91.042003}. Samples at the beginning of MCMC known as burn-in
period need to be discarded, because they have not explored the entire
parameter space and thus they are not guaranteed to be sampled from the
posterior. The termination condition can be set to the desired number of
samples, however, adjacent samples are usually correlated. We compute
autocorrelation time $\tau$ \cite{PhysRevD.91.042003}, and only select
effective samples: every $\tau$-th samples after burn-in, which can better
sample the posterior as they are independently chosen. 

The priors for mass and spin $(\chi)$ that we used for GW170817 are consistent with the
priors presented in Ref.~\cite{LIGOScientific:2019eut}, specifically, we
consider narrow and broad priors on masses and spins. Our choice of the narrow
prior is based on BNS observed in our Galaxy. We assume component masses of 
BNS follow uncorrelated Gaussian distribution with mean $1.33\,M_{\odot}$ and 
standard deviation $0.09\,M_{\odot}$ \cite{annurev-astro-081915-023322}, and 
employ the sorting convention $m_2 \leq m_1$. The choice of the upper limit of spin 
in this case is $\chi \leq 0.05$, which is identical to Ref.~\cite{LIGOScientific:2019eut} 
and motivated from observational data of spin of NS in binary systems that 
will merge in Hubble time. For the broad prior, the masses are uniformly distributed 
between $0.7\,M_{\odot}$ and $3.0\,M_{\odot}$. Again, following Ref.~\cite{LIGOScientific:2019eut}, 
we choose the prior  on $\chi$ in this case is uniformly distributed between 0 and 0.7.  
For GW190425, a high-mass BNS with total mass $\sim 3.4\,M_{\odot}$ and chirp mass
$\sim 1.44\,M_{\odot}$ \cite{Abbott:2020uma, gw190425}, we only employ the broad prior, 
identical to what was used for the broad prior runs for GW170817. Finally, the choice of low 
frequency cutoff usually depends on the sensitivity of the detectors. For GW170817 the 
low frequency cutoff is 23\,Hz \cite{LIGOScientific:2019eut}, and we reduce the low 
frequency cutoff of GW190425 to 19.4\,Hz \cite{Abbott:2020uma}, due to the improvement
of the sensitivity.

In the injection study discussed in Sec.~\ref{sec:stacking}, we inject
GW170817-like BNS with $(1.4, 1.4)\,M_{\odot}$ at different distances 40\,Mpc,
70\,Mpc, 100\,Mpc, 130\,Mpc, and 160\,Mpc to test how the results vary with
respect to signal-to-noise ratio. The $\tilde{\Lambda}$ and $\delta\tilde{\Lambda}$ for
injections are computed using the APR4\_EPP equation of the state, and we use TaylorF2 both 
for the injected waveform and the waveform used for recovery.

\section{Results}
\label{sec:results}
In the following we will test the accuracy of our new Bayes-factor approximation technique by applying 
it to GW170817 data and then comparing the results to those obtained in Ref.~\cite{LIGOScientific:2019eut} 
using the standard method. We will then carry forward this method to demonstrate the effect of combining 
the results from multiple events.

\subsection{Model selection for GW170817}
\label{sec:gw170817}
\begin{figure*}[ht]
\centering
\includegraphics[width=0.90\textwidth]{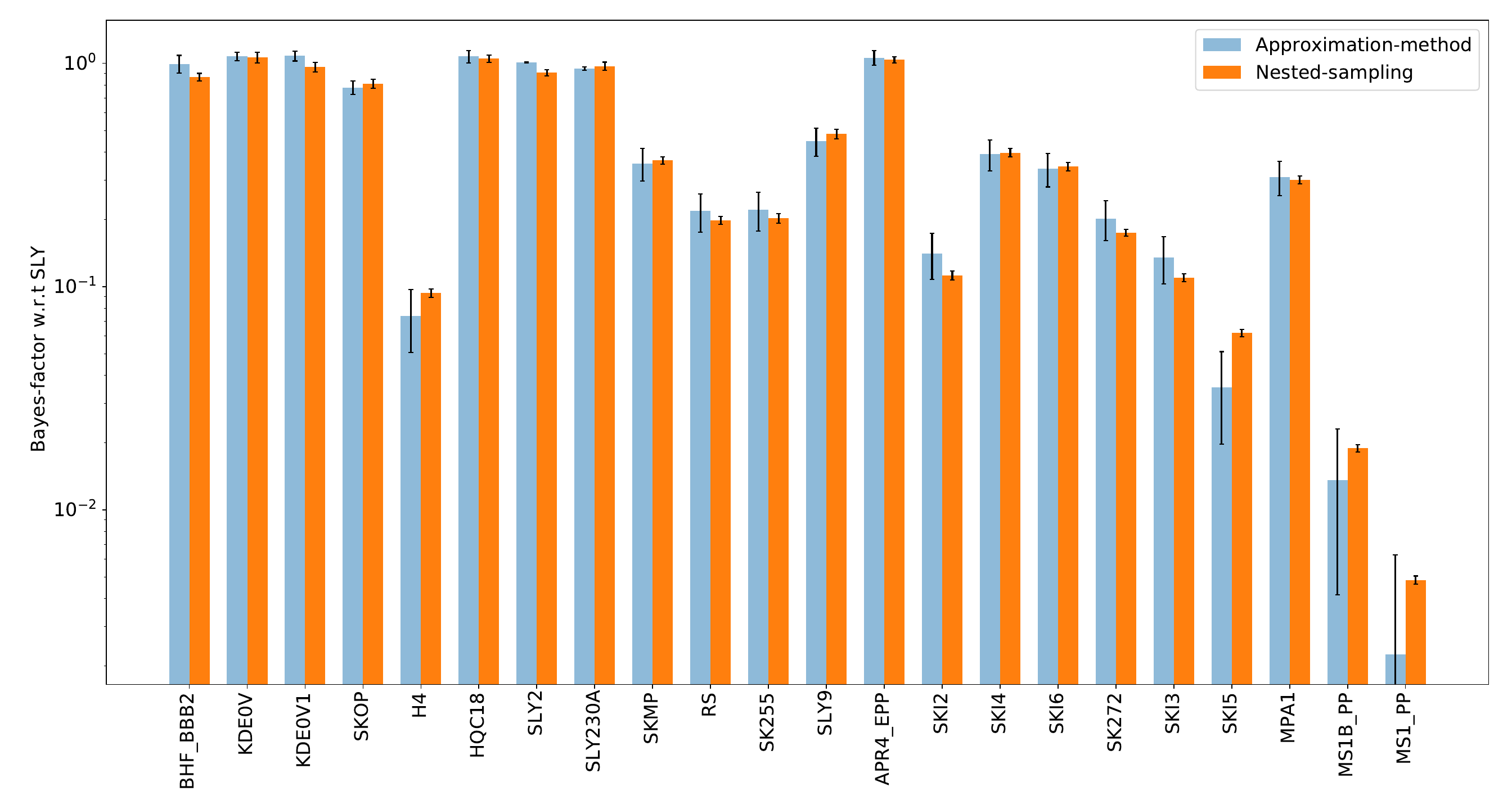}
\includegraphics[width=0.90\textwidth]{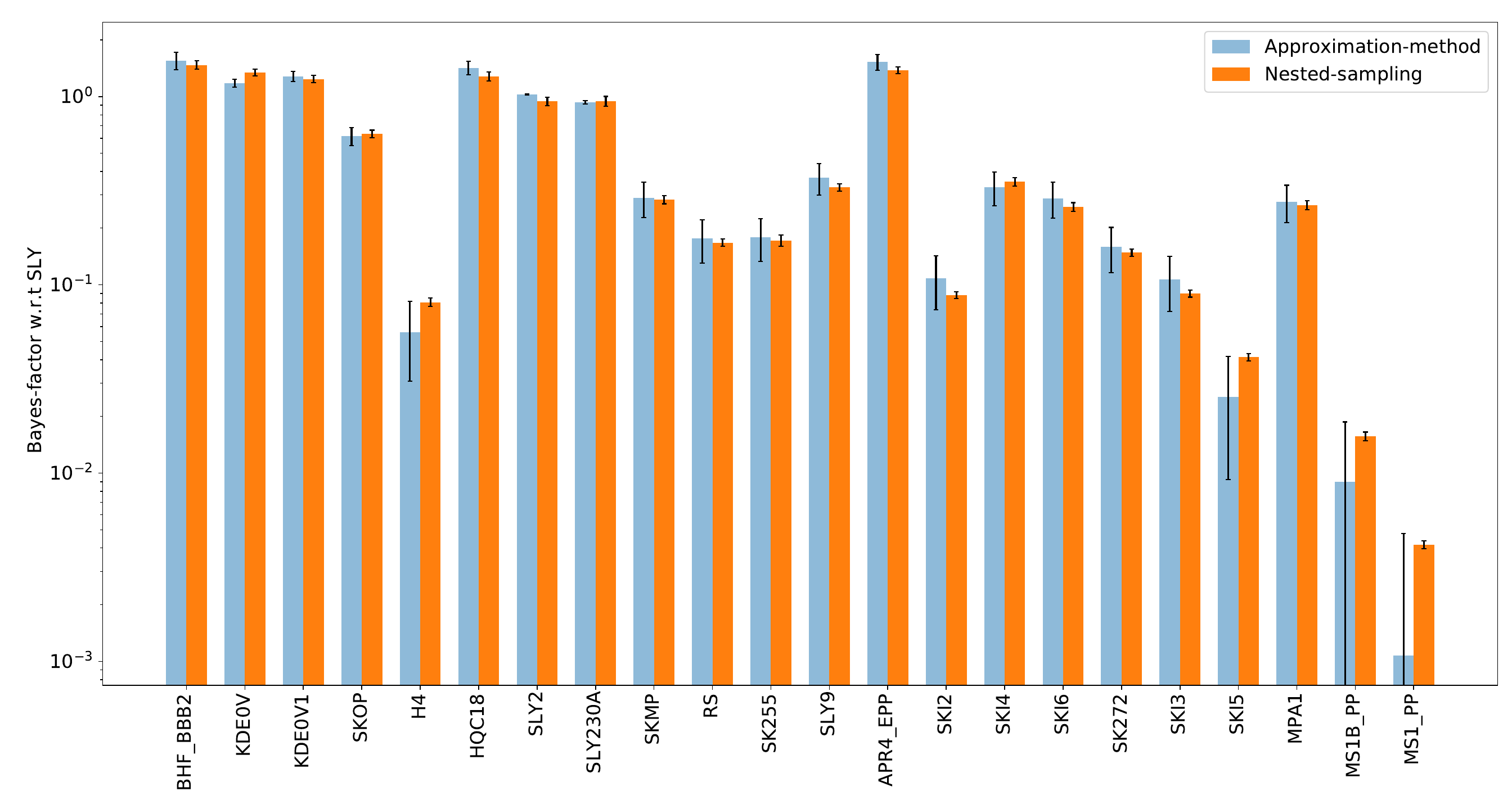}
\caption{\label{fig:evidence_ratio_post_sup}Comparison of Bayes factors
obtained with our approximate Bayes factor calculation scheme and the
\textsc{\texttt{LALInference\_nest}} nested sampling results
\cite{LIGOScientific:2019eut}, shown here for the narrow prior in the top
panel and the broad prior in the bottom panel. The Bayes factors are
computed with respect to the SLY equation of state model in both cases. We
show here results for the TaylorF2 waveform. The error-bars for the approximate 
Bayes factors are an estimate based on standard deviation (see main text).
The error-bars in the nested sampling method is obtained from 
\cite{LIGOScientific:2019eut}.
}
\end{figure*}

We first demonstrate the consistency of the values of the Bayes factor computed
using this approximation method with respect to the same quantities computed by exploring
the full parameter space. We conducted two Bayesian Markov chain Monte Carlo
simulations on the data surrounding the trigger time of GW170817. 
These two analyses involved the two aforementioned prior distribution of the source
parameters, a narrow prior and a broad prior of mass of spin distribution. 
Finally, as explained in Sec.~\ref{sec:approxevidencederivation}, we employ the
TaylorF2 waveform for convenience.  We chose to conduct our analysis on the set
of equation of state models used in \cite{LIGOScientific:2019eut} to facilitate
direct comparison. The result of this analysis is presented in Tables
\ref{tab:narrow-prior-bayesfactors} for the narrow prior case, and
\ref{tab:broad-prior-bayesfactors} for the broad prior case. The
Bayes factors in columns 3 and 4 are computed with respect to the SLY model for each of
the target \eos{} models named in the first column. We estimate the uncertainty
in the approximation method by the method of resampling (ten thousand times)
delineated in the preceding section. This is also shown in
Fig.~\ref{fig:evidence_ratio_post_sup}, where the Bayes factors are plotted as
vertical bars. The blue colored bars show the results for the approximation
method, and the orange bars show the corresponding Bayes factors for the nested
sampling runs as presented in \cite{LIGOScientific:2019eut}. The top panel
shows the results for the narrow prior case and the bottom panel shows the
same for the broad prior. The uncertainties, shown as error-bars, are multiplied 
by a factor of 2 as discussed above to accommodate biases in the uncertainty estimation
from KDE smoothing. Note that the uncertainties of the SLY2 and SLY230A models are 
extremely small because of their covariance with SLY, the reference model for these studies.

\begin{table*}[t]
\centering
\begin{tabular}{c c c c}
\hline
      EOS & $m_{\max}$ $(M_{\odot})$ & Bayes factor from Nest & Approx Bayes factor  \\
\hline
 BHF\_BBB2 &    1.922 &                  0.867 &               0.994 \\
    KDE0V &     1.96 &                  1.062 &               1.075 \\
   KDE0V1 &    1.969 &                  0.962 &               1.079 \\
     SKOP &    1.973 &                  0.811 &                0.78 \\
       H4 &    2.031 &                  0.094 &               0.074 \\
    HQC18 &    2.045 &                   1.05 &               1.074 \\
     SLY2 &    2.054 &                  0.908 &                1.01 \\
  SLY230A &    2.099 &                  0.972 &               0.947 \\
     SKMP &    2.107 &                  0.368 &               0.356 \\
       RS &    2.117 &                  0.198 &               0.218 \\
    SK255 &    2.144 &                  0.203 &                0.22 \\
     SLY9 &    2.156 &                  0.483 &               0.448 \\
 APR4\_EPP &    2.159 &                  1.037 &                1.06 \\
     SKI2 &    2.163 &                  0.112 &                0.14 \\
     SKI4 &     2.17 &                  0.398 &               0.392 \\
     SKI6 &     2.19 &                  0.345 &               0.337 \\
    SK272 &    2.232 &                  0.174 &               0.202 \\
     SKI3 &     2.24 &                   0.11 &               0.135 \\
     SKI5 &     2.24 &                  0.062 &               0.035 \\
     MPA1 &    2.469 &                  0.301 &               0.309 \\
  MS1B\_PP &    2.747 &                  0.019 &               0.014 \\
   MS1\_PP &    2.753 &                  0.005 &               0.002 \\
\hline
\end{tabular}
	\caption{Comparison of Bayes factors with respect to SLY \eos{} for the narrow prior. The results of the third column is calculated from \cite{LIGOScientific:2019eut}.}
  \label{tab:narrow-prior-bayesfactors}
\end{table*}

\begin{table*}[t]
\centering
\begin{tabular}{c c c c}
\hline
      EOS & $m_{\max}$ $(M_{\odot})$ & Bayes factor from Nest & Approx Bayes factor \\
\hline
\midrule
 BHF\_BBB2 &    1.922 &                   1.47 &               1.555 \\
    KDE0V &     1.96 &                  1.342 &               1.177 \\
   KDE0V1 &    1.969 &                  1.239 &               1.283 \\
     SKOP &    1.973 &                  0.634 &               0.618 \\
       H4 &    2.031 &                  0.081 &               0.056 \\
    HQC18 &    2.045 &                  1.278 &               1.422 \\
     SLY2 &    2.054 &                  0.945 &               1.028 \\
  SLY230A &    2.099 &                  0.945 &               0.932 \\
     SKMP &    2.107 &                  0.284 &                0.29 \\
       RS &    2.117 &                  0.167 &               0.176 \\
    SK255 &    2.144 &                  0.172 &               0.179 \\
     SLY9 &    2.156 &                  0.329 &                0.37 \\
 APR4\_EPP &    2.159 &                  1.382 &               1.526 \\
     SKI2 &    2.163 &                  0.088 &               0.108 \\
     SKI4 &     2.17 &                  0.352 &                0.33 \\
     SKI6 &     2.19 &                  0.259 &               0.288 \\
    SK272 &    2.232 &                  0.148 &               0.159 \\
     SKI3 &     2.24 &                   0.09 &               0.107 \\
     SKI5 &     2.24 &                  0.041 &               0.025 \\
     MPA1 &    2.469 &                  0.265 &               0.276 \\
  MS1B\_PP &    2.747 &                  0.016 &               0.009 \\
   MS1\_PP &    2.753 &                  0.004 &               0.001 \\
\hline
\end{tabular}
	\caption{Comparison of Bayes factors with respect to SLY \eos{} for the broad prior. The results of the third column is calculated from \cite{LIGOScientific:2019eut}.}
  \label{tab:broad-prior-bayesfactors}
\end{table*}

\begin{figure*}[ht]
\centering
\includegraphics[width=0.9\textwidth]{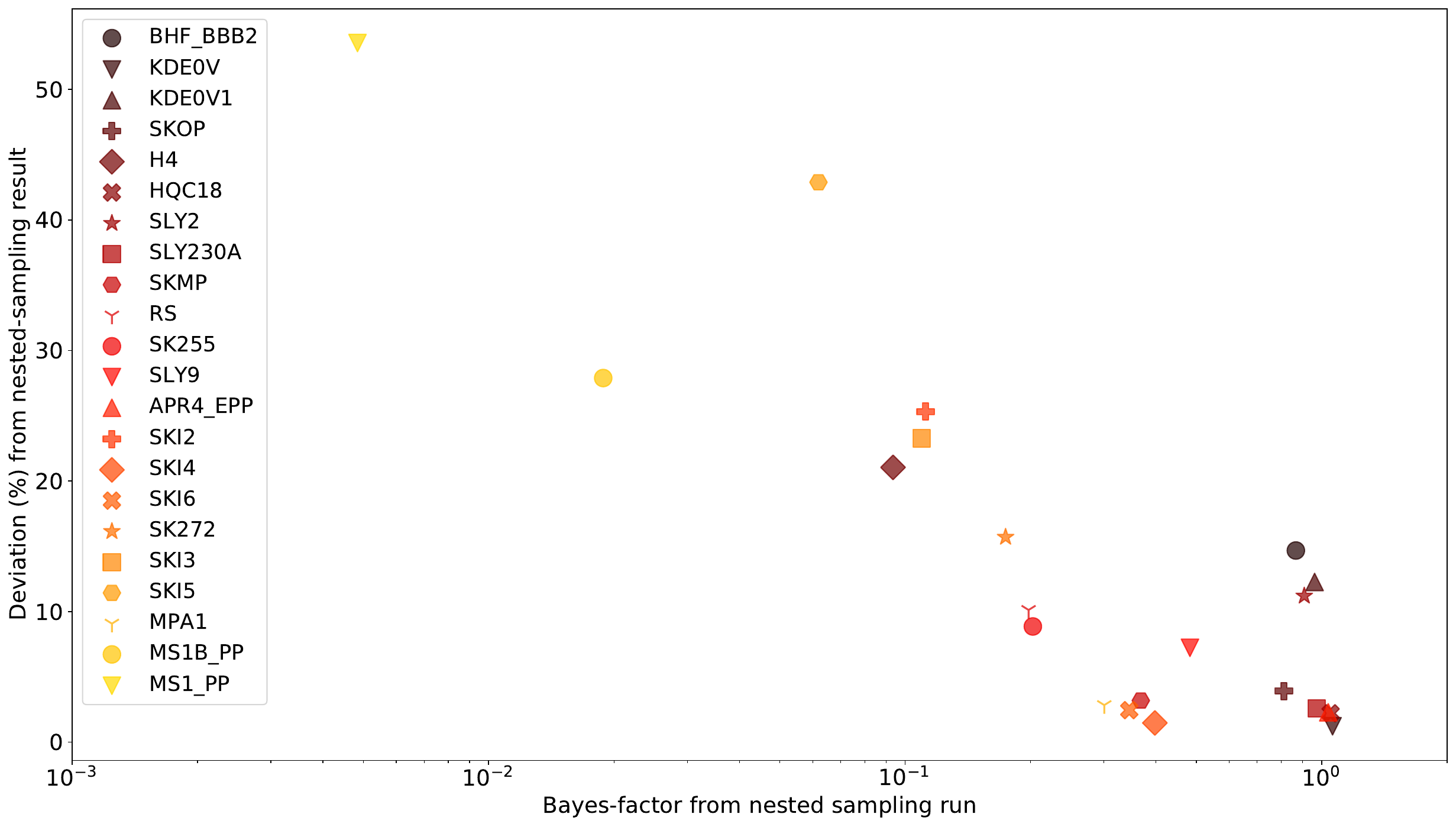}
\includegraphics[width=0.9\textwidth]{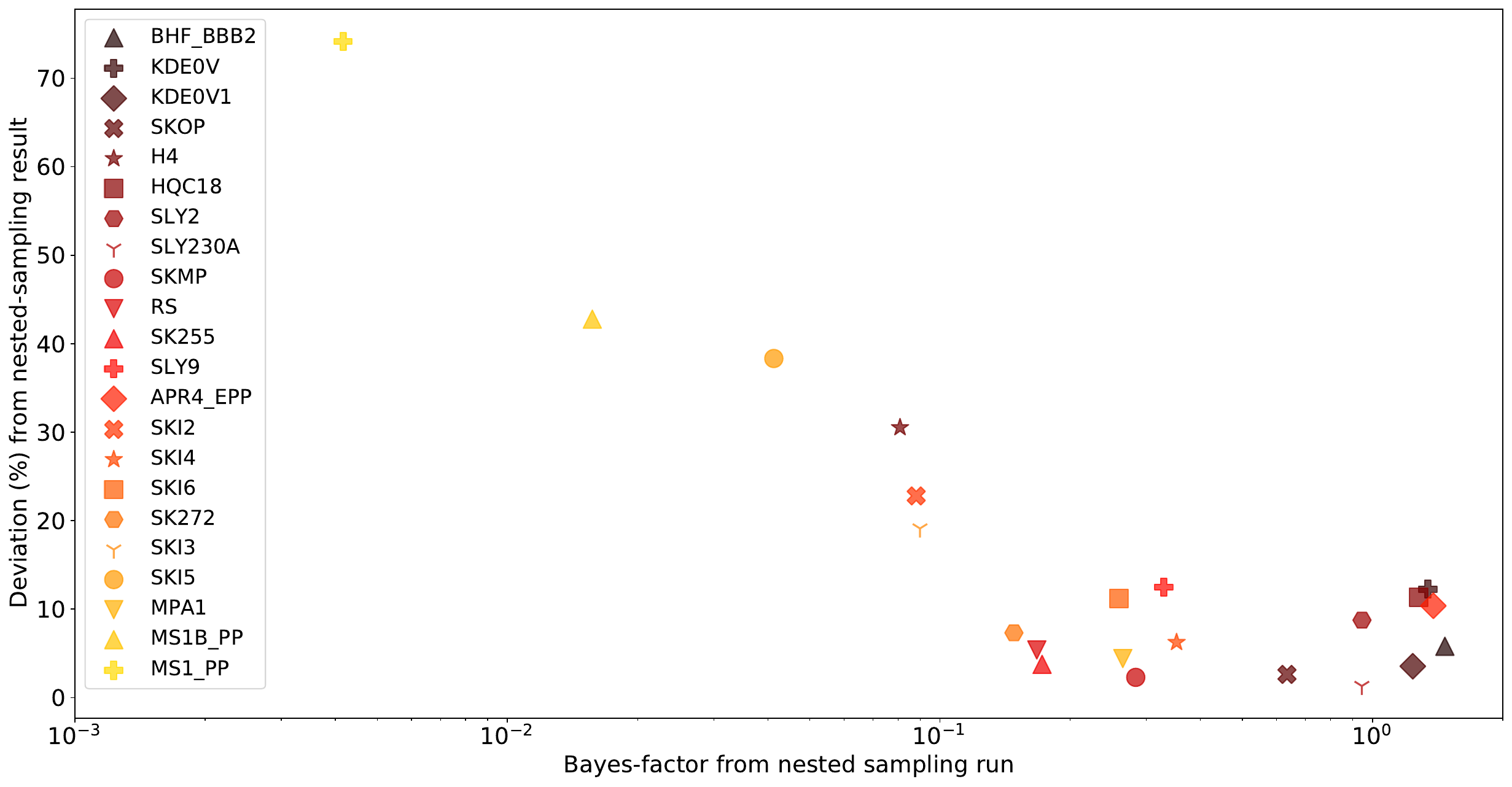}
\caption{\label{fig:evidence_ratio_deviation}Magnitude of deviation of Bayes factors
obtained with our approximate evidence calculation scheme from the
\textsc{\texttt{LALInference\_nest}} nested sampling results, shown here for the narrow prior 
in the top panel and the broad prior in the bottom panel. The Bayes factors are again computed 
with respect to the SLY equation of state model in both cases. We show here results for the 
TaylorF2 waveform. The deviation is the highest for the stiffest \eos{} which has the least 
evidence in the data. The method is more reliable when the model has good support from 
the data. The median deviation for the narrow (broad) prior is $10\% (11\%)$.}
\end{figure*}

\begin{figure*}[ht]
\centering
\includegraphics[width=0.49\textwidth]{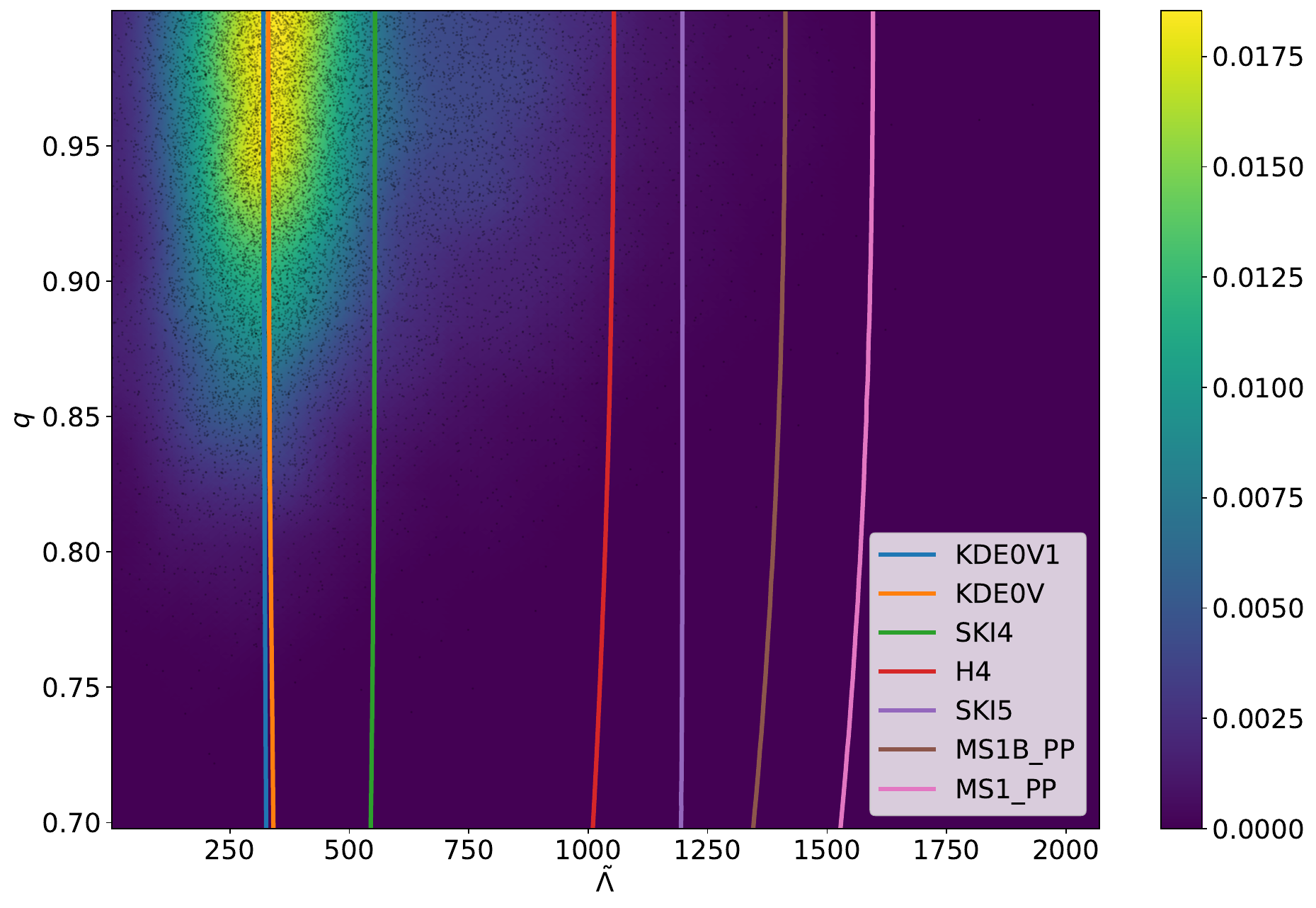}
\includegraphics[width=0.49\textwidth]{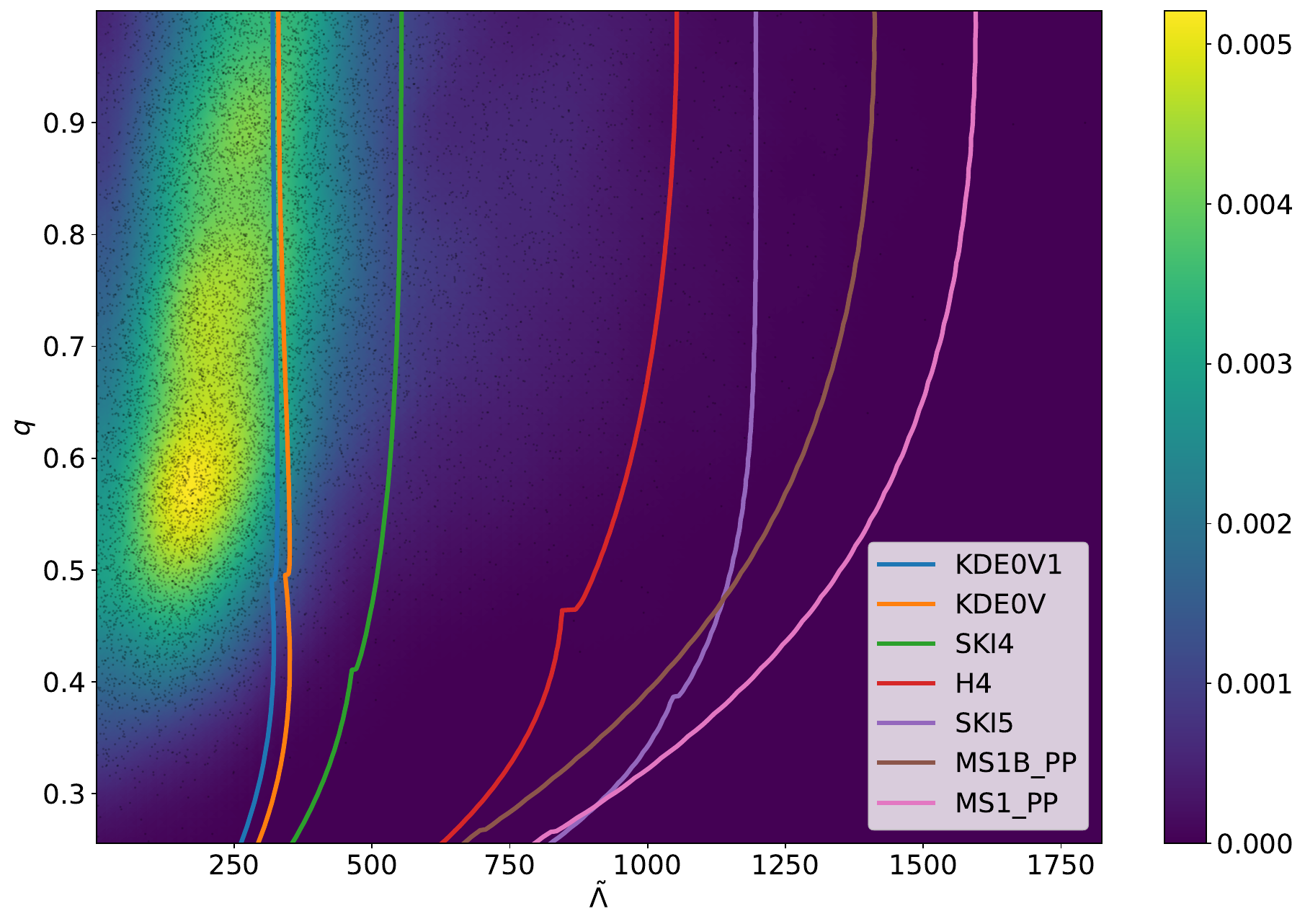}
\caption{\label{fig:posterior_support_narrow_broad} {\bf Left:} narrow prior: 
posterior distribution in $(\tilde{\Lambda}, q)$ and its KDE\@.
For comparison we show the \eos{} curves for various models in $(\tilde{\Lambda},
q)$. Note that the models that gave the highest deviation with respect to the nested
sampling results in Fig.~\ref{fig:evidence_ratio_deviation} are also the most
distant from the peak of the posterior distribution. 
{\bf Right:} broad prior case: we see the
identical relationship that for the \eos~models that gave the largest deviation
in the Bayes factor computation with respect to the nested sampling results, the
posterior support is the weakest. Note that the kinks in the various
\eos{} models in this plot are due to the fact that it extends to smaller values
of $q$, where one of the objects in the binary becomes more massive than the
maximum allowed NS mass. At this point we consider that object a
black hole and set the value of $\Lambda_1=0$. This leads to a sudden change in
the value of $\tilde{\Lambda}$ and hence creates these kinks.}
\end{figure*}

The results produced from the evidence approximation scheme are generally in
good agreement with those produced using the nested sampling method. Note,
however, that the Bayes factors obtained within the approximation scheme have
large fractional errors for equations of state predicting low evidences. This
is due to the intrinsically poor sampling in the regions of parameter space
that are least likely {\em a posteriori}. The number of samples produced at very large 
tidal deformabilities is dwarfed by the number of samples produced at more 
modest tidal deformabilities, because softer equations of state are more preferred 
by the GW170817 data.  This is why we provide the Bayes factors with 
respect to the SLY \eos{} as reference model for which the standard deviation of 
the evidence computation using the approximation method was relatively small, 
thus reducing the reference model contribution to the Bayes factor residuals. 
In Fig.~\ref{fig:evidence_ratio_deviation} we also show the deviation of the 
approximation method result with respect to the nested sampling method runs. 
The deviation is quantified as the absolute value of the difference between the 
mean value of the Bayes factor calculated by the two methods, divided by the 
mean value of the Bayes factor calculated by the nested sampling method. In 
the vertical axis of Fig.~\ref{fig:evidence_ratio_deviation} we show this deviation 
as percentage, and find that the deviation is highest for the stiffest \eos{} models 
for both the priors. This is the direct consequence of the sparseness of posterior 
samples in the region of the parameter space consistent with these \eos{} models. 
The kernel density estimation of the probability density is less reliable at these parts 
of the parameter space. This is further illustrated in 
Fig.~\ref{fig:posterior_support_narrow_broad} where we plot the $(\tilde{\Lambda},
q)$ posterior distribution for both the priors (narrow and broad) as a
scatter plot, along with their KDE as the heat-map. The $(\tilde{\Lambda}, q)$
values for some of \eos{} models are overlaid on top of the plots. It is
immediately evident that the \eos{} models which had the highest deviation in
Fig.~\ref{fig:evidence_ratio_deviation} are the ones that are most distant from
the peak of the posterior distribution, confirming the aforementioned argument
that the KDE is less reliable when number of underlying posteriors samples is very low.

An interesting observation can be made from the comparison plot in
Fig.~\ref{fig:evidence_ratio_post_sup} top panel. The two \eos{} models KDE0V and
KDE0V1 are shown to have Bayes factors of $1.075$ and $1.079$ respectively based
on the computation using the approximation method. However, the same two
\eos{} models have Bayes factors of $1.062$ and $0.962$ according to the nested
sampling based analysis. As a result of that the Bayes factors computed for
KDE0V1 using the two methods seem to be in disagreement. What is surprising is that 
these two models are very similar as shown in Fig.~\ref{fig:posterior_support_narrow_broad}, 
so one might expect that their Bayes factors as informed by the posterior sample distribution 
should also be similar. This seems to be the case with the approximation method, while the
nested-sampling results are less similar. We think this is happening due to the following
reason. For the approximation method, we compute the posterior samples just
once. Thus, the KDE obtained for the posterior samples is determined once and
for all for each \eos{} model. The Bayes factors for KDE0V and KDE0V1 are
computed using the same KDE, and since the two models are very similar the Bayes-factor 
values are also similar. 

The resampling technique used to compute the uncertainty captures the fluctuation in the
estimation of the KDE due to the finite number of samples. This is what is reported by the
error-bars in Fig.~\ref{fig:evidence_ratio_post_sup}. However, this technique does not allow
us to compute the uncertainty in the Bayes-factor from the posterior samples.
In the nested sampling method, however, the Bayes factor is computed for KDE0V and KDE0V1 separately
with different nested sampling parameter estimation runs. Thus, the underlying posterior distributions 
used for computation of the Bayes-factor for these two models are themselves different. The resulting 
Bayes factor is therefore  affected by the inherent variance of the nested sampling runs. The error-bars 
of the Bayes factors computed from the nested sampling run for the two \eos{} overlaps, suggesting 
that the true Bayes factors of these two  models are indeed very close to each other.

Finally, we discuss about the binary black hole model in our approximation
scheme. In Ref. \cite{LIGOScientific:2019eut} the authors have used a binary
black hole model as the reference model for the computation of the
Bayes factors. One of the shortcomings of the approximation method is that it
does not give reliable results for models around part of the parameter space
that does not have large posterior support.
Moreover, the KDE computation may be less reliable near the
boundaries of the parameter space, where posterior sample points consistent
with a binary black hole system will be located, leading to large biases in
computation of the evidence integral. With this in mind we avoided using binary
black hole as a reference \eos{} in this work, and have selected the SLY \eos{} as
the reference, as it was one of the \eos{} that has small uncertainty in the
value of the evidence integral.

A major advantage of the approximation scheme, shown here to reliably reproduce
the evidences calculated in \texttt{LALInference\_nest}, is that it allows us
to compute the evidence in a fraction of the time taken compared to a full
nested sampling run using \texttt{LALInference\_nest}. The code
to calculate this \cite{gwxtreme-pypi}, as well as demonstrations of how to
use the package, are released along with this work in \cite{gwxtreme-doc} .

\subsection{Stacking of multiple events}
\label{sec:stacking}
Assuming that the equation of state of NS matter is unique, it should be 
possible to combine information from multiple gravitational wave detections 
to produce joint inference on the \eos{} models. Multiple studies have been 
conducted to this end to get joint constraints \cite{DelPozzo:2013ala, 2015PhRvD..91d3002L}.
In this approximation scheme, multiple detection of gravitational wave events 
can be incorporated using stacking to make joint inferences on the Bayes-factor 
between the various models of equation of state. To do so, we simply incorporate 
this by computing the products of the evidences for the various models.
\begin{equation}
\label{eq:stacking}
\mathcal{Z}_{E}(d_1,\ldots,d_N) = \prod\limits_{i=1}^{N} \mathcal{Z}_{E}(d_i)\,,
\end{equation}
where $\mathcal{Z}_{E}(d_i)$ is the evidence of the equation of state model ${E}$ for event $i$, and $N$
is the number of detected gravitational wave events. Thus, the
joint Bayes factor for $N$ events is given by
\begin{equation}
\label{sec:jointbf}
{\rm BF}^{E_1}_{E_2}({\rm joint}) = \frac{\mathcal{Z}_{E_1}(d_1,\ldots,d_N)}{\mathcal{Z}_{E_2}(d_1,\ldots,d_N)}\,.
\end{equation}

The uncertainty in the joint Bayes factor can be estimated using multiple
techniques. In this work we have used the same bootstrapping technique 
described in Sec.~\ref{sec:approxevidencederivation}. For a given pair of 
equations of state, we resample the Bayes factors for all individual events 
ten thousand times to obtain a distribution of the joint Bayes factor. The
variance of this quantity gives us the estimate of the uncertainty. Following
the same procedure we augment the uncertainties of the Bayes factors by a
factor of 2 to take into account the effect of oversmoothing of the posterior
samples and hence render a conservative estimate.

\begin{figure*}[h]
\centering
\includegraphics[width=1.0\textwidth]{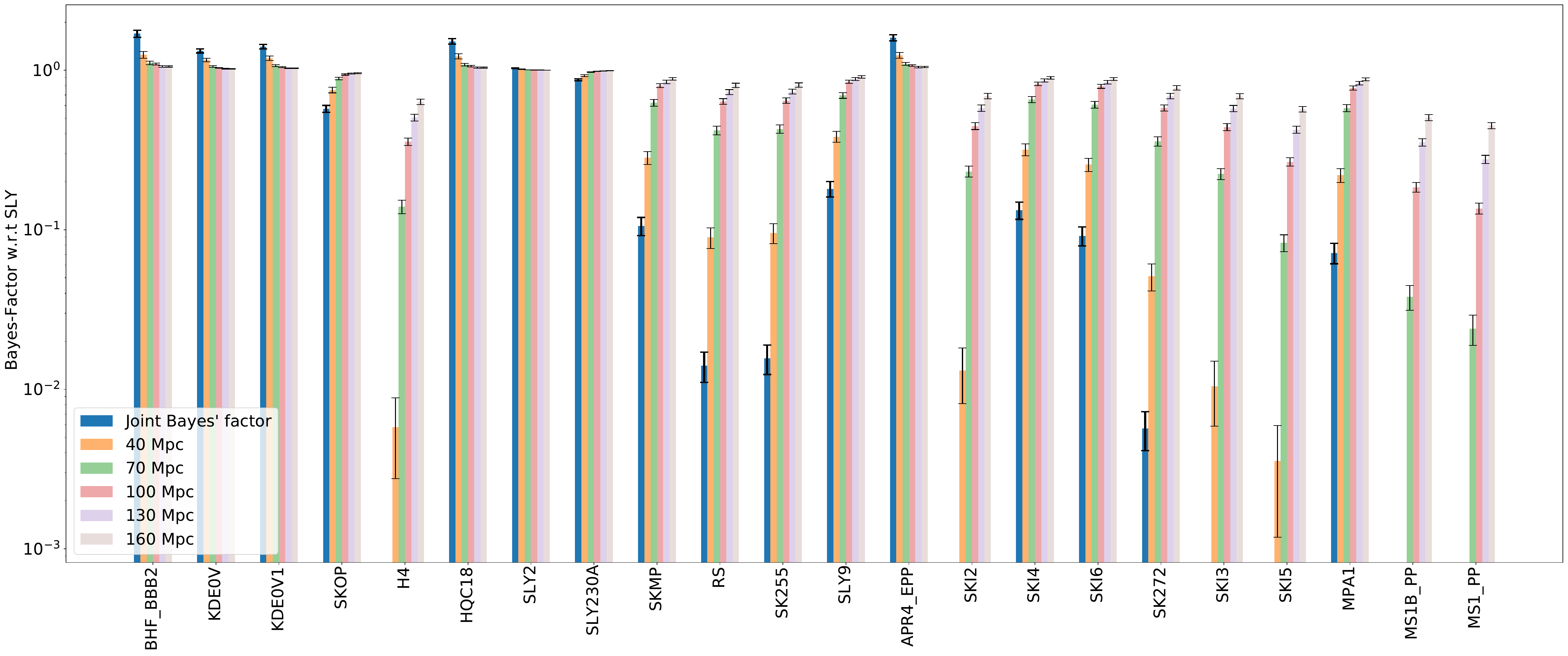}
\caption{\label{fig:bayes-factor-stacking} Bayes-factors computed for the various \eos{} models for the injected signals at 40 Mpc, 70 Mpc,
100 Mpc, 130 Mpc, and 160 Mpc. 
The blue bars shows the joint Bayes-factors computed using Eq. \ref{sec:jointbf}, which is largely determined by the strongest source. A 
Bayes-factor threshold of $10^{-3}$ is applied to exclude cases that have very small posterior support.}
\end{figure*}

To test the performance of this stacking method, we conducted a simulation study.
We injected the gravitational wave signal of a binary NS system with component 
masses $(1.4, 1.4)\, M_{\odot}$ in simulated Gaussian noise. The tidal deformability
of the system was chosen to be consistent with the APR4\_EPP equation of
state. We injected five gravitational waveforms at 40\,Mpc, 70\,Mpc, 100\,Mpc, 130\,Mpc and 160\,Mpc respectively.
For each event we conducted Bayesian MCMC using \texttt{LALInference MCMC} along with the implementation of the  
modification of the waveform termination condition discussed earlier. 
We chose the same broad prior for masses and spins that was described in Sec.~\ref{sec:approxevidencederivation}.
The result of the study is presented in Fig.~\ref{fig:bayes-factor-stacking}. We apply a Bayes-factor threshold of $10^{-3}$ in 
our analysis to exclude some of the stiffest \eos{} from the result. This is necessary because for these models the $(q, \tilde{\Lambda})$ 
posterior samples are poorly sampled by the parameter estimation run, resulting in KDE estimation that is not reliable. The choice of
this threshold is motivated by the fact that this was the smallest value of the Bayes-factor for the study using GW170817 (for MS1\_PP). 
This issue can be alleviated by increasing the number of MCMC chains, however for the purpose of demonstration
of this method this is unnecessary. 

We expected to find that with increasing distance of the injected source the Bayes-factor 
to converge towards one. When we analyzed the APR4\_EPP model itself we found the Bayes-factor with respect to the 
SLY \eos{} to be 1.24, 1.09, 1.07, 1.05, and 1.05 for the above mentioned injected distances respectively. 
Using the stacking method described above, this led us to a joint Bayes-factor of 1.6 between the APR4\_EPP
and SLY models. For a disfavored model like SK272 the discrimination ability of the method deteriorates rapidly with 
lowering of the signal strength. While at 40 Mpc the Bayes-factor for SK272 with respect to SLY is $0.05$, at 70 Mpc, 
100 Mpc, 130 Mpc, and 160 Mpc the Bayes-factors are 0.36, 0.58, 0.69, 0.78 respectively, giving a joint Bayes-factor of 
$5.69\times 10^{-3}$. 
In Fig.~\ref{fig:bayes-factor-stacking} we find that, as expected, the height of the bars are tending towards the asymptotic 
value of 1.0 as we increase the distance of the source. Note that the heights are more uniform between the different models
for the farthest sources. This is due to the fact that weaker signals leave less of an imprint of the tidal deformability in the 
data to help in discerning between different models. The joint Bayes-factor will be mostly determined by the strongest sources. 
However, one should also keep in mind the $r^3$ distribution of sources. In practice we expect to see more sources at larger 
distances and fewer sources at closer distances. The combined effect of a large number of weak sources and a small number of 
strong sources will collectively improve our understanding of the NS \eos.

\subsection{Joint Bayes factor from GW170817 and GW190425 using stacking}
\label{sec:stacking_real}
\begin{figure*}[h]
\centering
\includegraphics[width=1.0\textwidth]{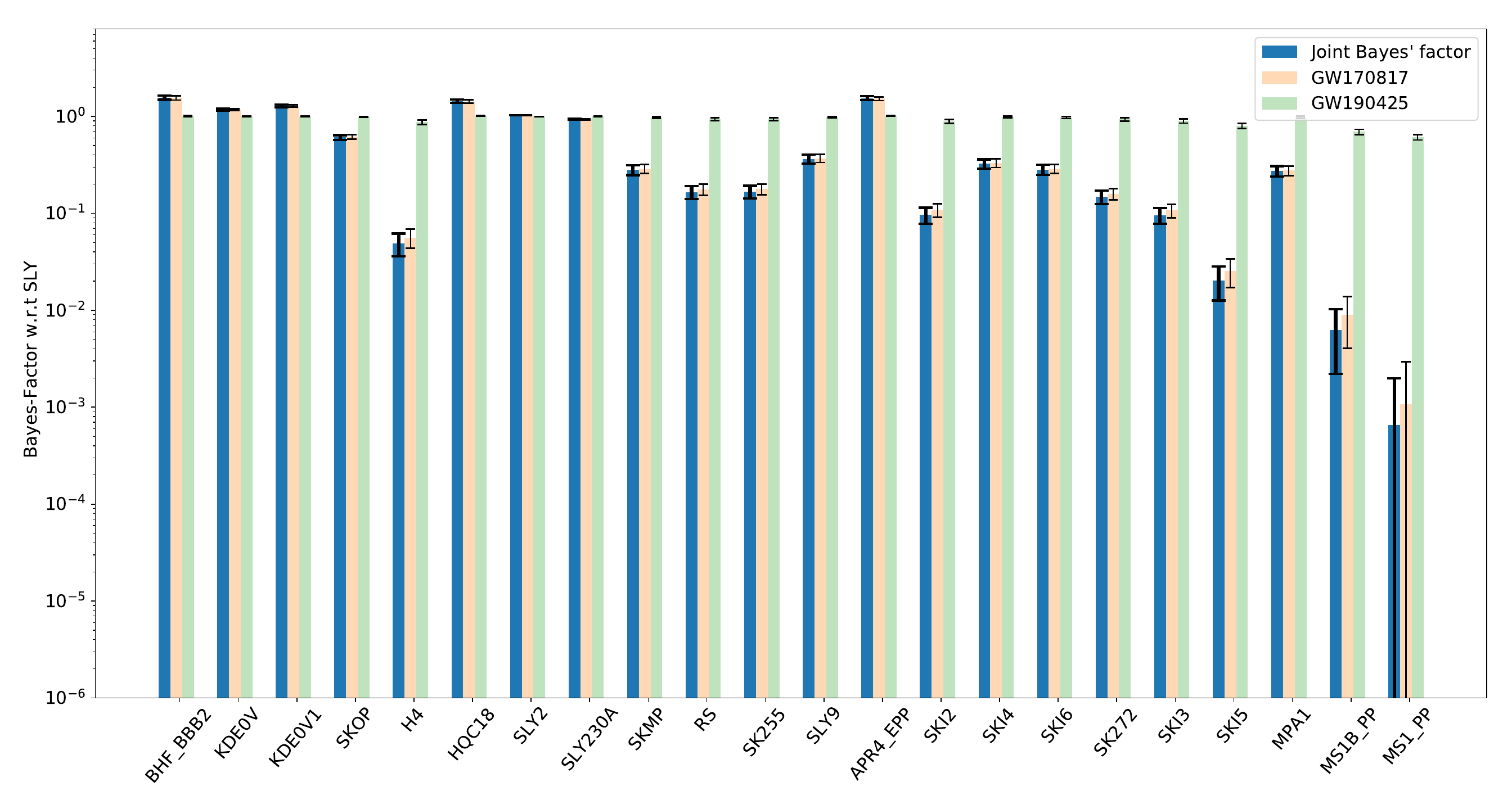}
\caption{\label{fig:stack-gw170817-gw190425} Bayes factor of various
\eos{} models with respect to SLY model for GW170817, GW190425, and their combination by
evidence stacking. Note that the blue bars (joint Bayes factor) and the orange
bars (Bayes factor from GW170817 data) are very similar to each other
indicating that most of the information in discerning between the different
models comes from the data of GW170817. The green bars for GW190425 are adding
very little information as can also be seen from the fact that their variation
in height across the various models is very small.}
\end{figure*}

In Sec.~\ref{sec:stacking} we introduced the method of stacking in the evidence
approximation technique to combine data from multiple BNS coalescence events 
to get a joint Bayes factor between pairs of \eos~models. In this section we use 
this method to combine the gravitational-wave data from GW170817 and GW190425, 
the two BNS coalescences that the LIGO-Virgo interferometers have detected. What 
makes GW190425 especially interesting is that the heavier object in the binary is 
estimated to be around $1.60\,M_\odot$ to $2.52\,M_{\odot}$ (if we apply a broad 
prior of the object as mentioned in this work) \cite{Abbott:2020uma}. The upper-limit 
of the mass of this object is at the edge of maximum NS mass of some \eos{} models. 
Unfortunately, the luminosity distance of this event is $\sim 4$ times greater than the 
luminosity distance of GW170817. Thus, the strength of the gravitational wave from this 
event is much weaker across the entire frequency band. This reduction in signal 
strength, especially in the high frequency regime, severely affects our ability to infer on 
the tidal deformability and hence the NS \eos. Thus, we do not expect a very large effect 
of including the data from GW190425 in the computation of the Bayes factor between the 
various \eos{} models.  We conducted Bayesian MCMC parameter estimation run on the data from 
the interferometers using the specifications detailed in \ref{sec:pe-gw190425}. The 
posterior samples from this run were then used to compute the Bayes factors against the 
SLY \eos{} using the approximation method. We then combine the Bayes factors for the 
various models with respect to SLY with the same computed for GW170817 using the method 
of evidence-stacking discussed in Sec.~\ref{sec:stacking}. We show in 
Fig.~\ref{fig:stack-gw170817-gw190425} the result of this analysis.

\begin{figure*}[h]
\centering
\includegraphics[width=1.0\textwidth]{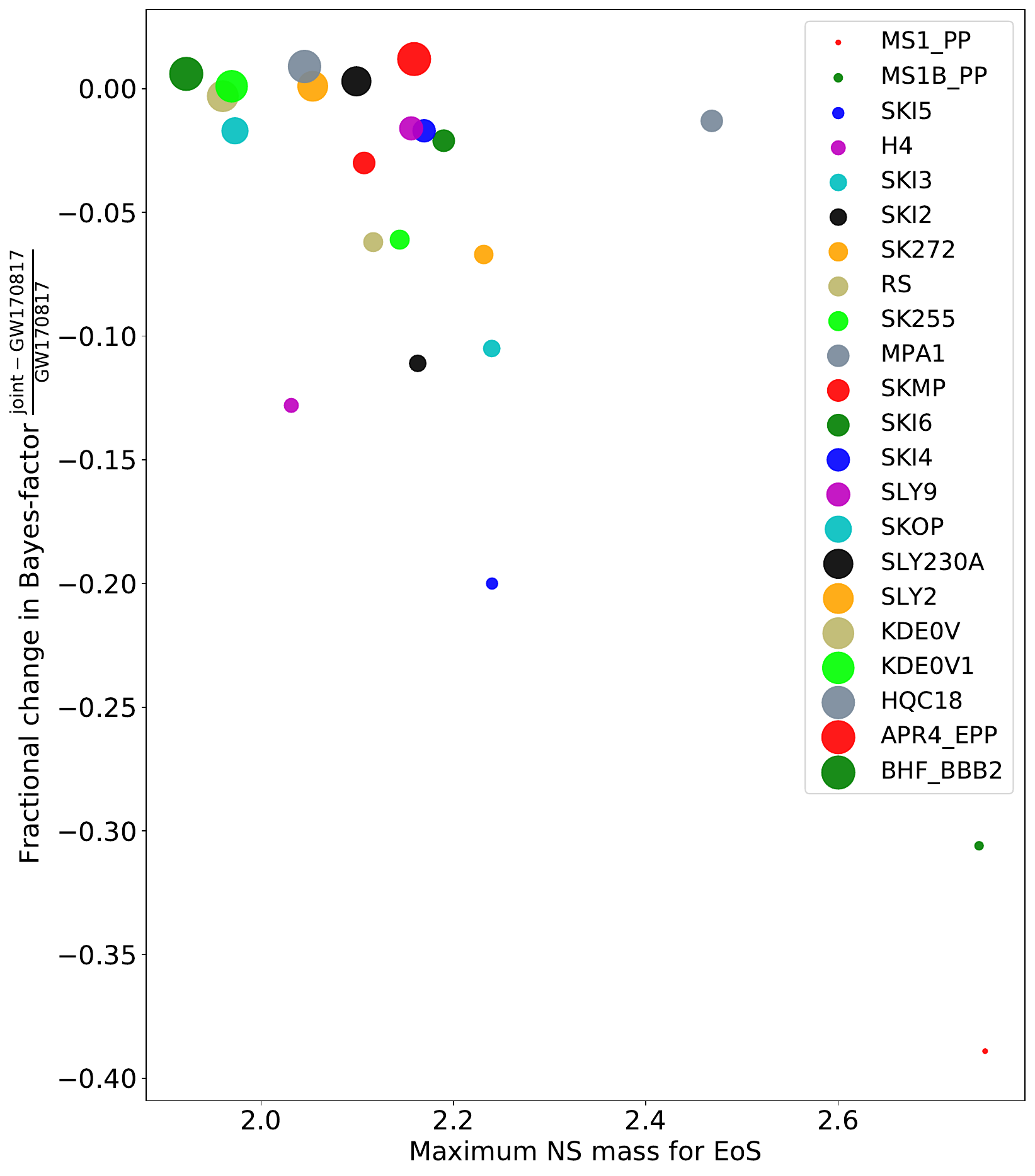}
\caption{\label{fig:stack-gw170817-gw190425-change} Change in Bayes factor due
to inclusion of GW190425 for different \eos{} models. The area of the bubbles are proportional
to the respective joint Bayes factor values. We note that the models for which the Bayes factor 
was higher for GW170817 tend to an increased Bayes-factor upon inclusion of the GW190425 
data, whereas models which have lower Bayes-factor for GW170817 diminish further upon 
inclusion of the GW190425 data. However, it should be noted that this effect is very small, 
given the weakness of the signal from GW190425, as evidenced by the consistency between 
the bars of Bayes-factors of GW170817 and the joint Bayes-factor in Fig. \ref{fig:stack-gw170817-gw190425}}
\end{figure*}

\begin{table*}[t]
\centering
\begin{tabular}{c c c c c c}
\hline
      EOS & $m_{\max}$ & BF(GW170817) & BF(GW190425) & BF(joint) & Fractional change \\
\hline
 BHF\_BBB2 &    1.922 &     1.555 &     1.006 &    1.564 &             0.006 \\
    KDE0V &     1.96 &     1.177 &     0.997 &    1.174 &            -0.003 \\
   KDE0V1 &    1.969 &     1.283 &     1.001 &    1.285 &             0.001 \\
     SKOP &    1.973 &     0.618 &     0.983 &    0.607 &            -0.017 \\
       H4 &    2.031 &     0.056 &     0.872 &    0.049 &            -0.128 \\
    HQC18 &    2.045 &     1.422 &     1.009 &    1.436 &             0.009 \\
     SLY2 &    2.054 &     1.028 &     1.001 &    1.029 &             0.001 \\
  SLY230A &    2.099 &     0.932 &     1.003 &    0.935 &             0.003 \\
     SKMP &    2.107 &      0.29 &      0.97 &    0.281 &             -0.03 \\
       RS &    2.117 &     0.176 &     0.938 &    0.166 &            -0.062 \\
    SK255 &    2.144 &     0.179 &     0.939 &    0.168 &            -0.061 \\
     SLY9 &    2.156 &      0.37 &     0.984 &    0.364 &            -0.016 \\
 APR4\_EPP &    2.159 &     1.526 &     1.012 &    1.544 &             0.012 \\
     SKI2 &    2.163 &     0.108 &     0.889 &    0.096 &            -0.111 \\
     SKI4 &     2.17 &      0.33 &     0.983 &    0.325 &            -0.017 \\
     SKI6 &     2.19 &     0.288 &     0.979 &    0.282 &            -0.021 \\
    SK272 &    2.232 &     0.159 &     0.933 &    0.148 &            -0.067 \\
     SKI3 &     2.24 &     0.107 &     0.895 &    0.096 &            -0.105 \\
     SKI5 &     2.24 &     0.025 &       0.8 &     0.02 &              -0.2 \\
     MPA1 &    2.469 &     0.276 &     0.987 &    0.273 &            -0.013 \\
  MS1B\_PP &    2.747 &     0.009 &     0.694 &    0.006 &            -0.306 \\
   MS1\_PP &    2.753 &     0.001 &     0.611 &    0.001 &            -0.389 \\
\hline
\end{tabular}
\caption{Bayes factors for various \eos{} models computed for GW170817, GW190425
and their combined result. The last column shows the fractional change in the
Bayes factor for model when evidence GW190425 is stacked with that of GW170817:
$[{\rm BF(joint)} - {\rm BF(GW170817)}]/{\rm BF(GW170817)}$. We note a trend that
this fractional change in the value of the Bayes factor increases with
increasing stiffness of the \eos{} model. However, we also notice that there is a
direct correlation between the increase in this fractional change with value of
the joint Bayes factor.} \label{tab:fractional-change-bayesfactors}
\end{table*}
In Table \ref{tab:fractional-change-bayesfactors} we present the result of the
stacking for all the \eos{} models. In this table we show the result of the broad prior 
only, which is consistently employed for both events. The third column shows the Bayes 
factor computed using the GW170817 data only, in the fourth column we show the 
same for GW190425. In the fifth column we show the stacked result. The last column 
shows the fractional change in the Bayes factor due to the inclusion of the second event, 
GW190425. This is defined as:
\begin{equation}
\label{eq:approximationEquation}
\frac{{\rm BF(joint)} - {\rm BF(GW170817)}}{\rm BF(GW170817)}\,.
\end{equation}
We find that models which have large Bayes-factor in GW170817 data slightly gain in their 
Bayes-factor value upon inclusion of the data from GW190425. The gain is around a percent
or less. Models that have low Bayes-factors with respect to the SLY \eos{} as evidenced from GW170817
data, further diminishes in value upon inclusion of GW190425. This lowering of the Bayes-factor
can be substantial for some of the stiffest models like MS1.
We notice a slight trend of higher impact of using the second event for
stiffer \eos{} models (Fig.~\ref{fig:stack-gw170817-gw190425-change}). Note that
part of the reason why the Bayes factor changes more for such models could be
because of the greater random fluctuation in the individual Bayes factors for
them. However, we find that the change is biased toward lower values of the
Bayes factors upon stacking the events (hence the negative values in the Fig. \ref{fig:stack-gw170817-gw190425-change}). 
If the random fluctuation was the sole reason for this change, this bias should not be present. This provides us an
indication, as far as Bayes factor is concerned, that the stiffer models are
aggregating more information from GW190425. 
However, it should also be mentioned that the models for which
the addition of GW190425 makes a larger impact are also the ones that
have already been found to be highly disfavored by GW170817. In fact the 
values of joint Bayes-factor are very consistent with the Bayes-factor obtained
from GW170817 data, which is in line with the observation made in the injection
study in Sec. \ref{sec:stacking}. Therefore, to gather more insightful results on the 
\eos{} models we will need either stronger signals than GW190425, or a larger 
number of events.

\section{Discussions}
\label{sec:discussions}
The Bayes factor approximation method presented here is designed to help comparisons
between a large number of equation of state models for NS matter. The crucial 
element of this technique is that one only needs a single instance of an \eos{} agnostic 
parameter estimation run conducted with the appropriate priors described above. The 
approximate Bayes factor between any two arbitrary models can then be computed very 
rapidly thereafter. The accuracy of the method is reasonably good and continues to perform 
well even after stacking multiple events as documented in the studies conducted in
Secs.~\ref{sec:gw170817} and~\ref{sec:stacking}. We make this method available for
public use in the \texttt{GWXtreme} package \cite{gwxtreme-pypi}. Detailed
information on using this package is provided in the documentation of the
package \cite{gwxtreme-doc}. The results of this study, including the posterior samples from 
the parameter estimation analysis that are required for the approximation method are publicly
released along with this work \cite{dataset}.

The method can be further extended to estimate parameters describing equation of state models, 
such as piecewise polytrope or spectral models.  To do so, the evidence is interpreted as the marginalized 
likelihood $\mathcal{L}(d \mid \vec{\Gamma})$ where $\vec{\Gamma}$ are the parameters characterizing 
the model.  One can then impose a prior on these parameters and then the posterior distribution 
$p(\vec{\Gamma} \mid d)$ can be computed.  Evaluating this involves computing the evidence integral 
in Eq.~ \ref{e:1dlineintkde}.  With $N$ events we simply need to compute $N$ such 1D integrals.  This idea
is analogous to the method presented in \cite{2015PhRvD..91d3002L}, where the evidence integral is carried
out in $(m_1, m_2, \tilde{\Lambda})$ space. Under this approximation scheme we compute a 1D line integral 
instead. This will provide a joint posterior distribution of the piecewise polytrope parameters or the spectral 
parameters.

There are however three caveats that we would like
the users of the package to be aware of.  First, this method employs kernel
density estimation to conduct the integration for the computation of the
Bayes factor. The KDE over-smooths the underlying distribution of the sample,
and as a result biases the estimation of the uncertainty. The uncertainty
measurement technique used in this work is not capable of quantitatively taking
into consideration of this effect, and as a result we only provide an approximate error 
estimate. In future work we will provide a more quantitative handle
on the uncertainty.
Second, in the case of GW170817 the lack of samples at low tidal
deformability results in Bayes factor computation with large statistical
fluctuations when extremely soft models are used as the reference \eos. It is
generally safe to use a reference \eos{} that has good overlap with the posterior
distribution (hence the choice of SLY in this work). We encourage the use of
SLY (or similar) \eos{} as the reference model for the computation of Bayes factor
for any \eos{} that has not been covered in this analysis.  Finally, the choice
of the uniform priors in $\tilde{\Lambda}$ and $\delta\tilde{\Lambda}$ results
in some negative values in tidal deformabilities of the individual stars. This
will be a problem for any waveform generator that requires the individual tidal
deformabilities to be positive. In this work we chose to use the TaylorF2
waveform since only $\tilde{\Lambda}$ is required to generate the waveform. For
other waveforms we will have to switch back to a uniform prior in individual
tidal deformabilities. However, at that point samples would need to be
reweighted such that the approximation in Eq.~(\ref{e:1dlineint}) works. This
will be included in future release of the \texttt{GWXtreme}.

\section*{Acknowledgement}
The authors will like to thank Reed Essick for meticulously reading through the
manuscript and reviewing the code. The authors will also like to acknowledge
Katerina Chatziioannou for reviewing the posterior samples and the parameter
estimation analysis that was run to generate them. The authors will also like to
thank Tim Dietrich for his valuable suggestion to improve the scientific content
of the paper, and for conducting the internal review of the article. S.G., X.L., and 
J.C. will like to acknowledge NSF grant NSF PHY-1912649 that supported this 
work. Large fraction of the analysis of the data was performed on the Nemo 
cluster at the Leonard E. Parker Center for Gravitation, Cosmology and 
Astrophysics at the University of Wisconsin-Milwaukee, CIT cluster at Caltech, 
LHO cluster at the Hanford LIGO Observatory, and the LLO cluster at the 
Livingston LIGO Observatory operated by the LIGO Lab, supported by the NSF 
Grants PHY-1626190, PHY-1700765, PHY-0757058 and PHY-0823459.

This research has made use of data, software and/or web tools obtained from 
the Gravitational Wave Open Science Center (https://www.gw-openscience.org/), 
a service of LIGO Laboratory, the LIGO Scientific Collaboration and the Virgo 
Collaboration. LIGO Laboratory and Advanced LIGO are funded by the United 
States National Science Foundation (NSF) as well as the Science and Technology 
Facilities Council (STFC) of the United Kingdom, the Max-Planck-Society (MPS), 
and the State of Niedersachsen/Germany for support of the construction of 
Advanced LIGO and construction and operation of the GEO600 detector. 
Additional support for Advanced LIGO was provided by the Australian Research 
Council. Virgo is funded, through the European Gravitational Observatory (EGO), 
by the French Centre National de Recherche Scientifique (CNRS), the Italian Istituto 
Nazionale di Fisica Nucleare (INFN) and the Dutch Nikhef, with contributions by 
institutions from Belgium, Germany, Greece, Hungary, Ireland, Japan, Monaco, 
Poland, Portugal, Spain.

This material is based upon work supported by NSF’s LIGO Laboratory which is a 
major facility fully funded by the National Science Foundation.

\section*{Appendix: ESTIMATION OF UNCERTAINTY SCALING-FACTOR}
The KDE resampling method for estimating the uncertainty discussed in 
Sec.~\ref{sec:approxevidencederivation} cannot account for potential 
systematic bias inherent in the KDE approximation. We account for this 
by applying a correction factor of 2 to the estimated standard deviation. 
In the following we explain how this factor is chosen.

First, we set up a probability density function (PDF) 
$p_\mathrm{ex}(q, \tilde{\Lambda})$ that is intended to be an example 
of the PDF used in Eq.~\ref{e:1dlineint}, representative for events our method might be applied 
to. One option would be to use the KDE obtained from our GW170817 posterior sample 
distribution as PDF. However, the sharpness of features would be limited by the KDE
bandwidth. In our approximation method we employ Scott's bandwidth given by $N^{-1/6}$, 
where $N$ is the number of posterior samples. To obtain an example PDF with somewhat 
sharper features, we duplicate each posterior sample and then create the KDE.

Since the example posterior is provided as a PDF instead of a sample distribution, 
the corresponding evidence for a given equation of state can be computed exactly. 
We compute the exact Bayes factors with respect to the SLY model, $B_e$, for a 
representative subset of equations of state, consisting of 
APR4\_EPP, H4, KDE0V, SKI2, SKI5, MS1B\_PP, and MS1\_PP, 
covering a wide range of stiffness.

To simulate a posterior sample distribution as obtained from a parameter estimation run, 
we draw the same number of samples from $p_\mathrm{ex}$ as contained in our real posterior 
sample distribution for GW170817. We do this repeatedly, creating 500 example posterior
sample distributions. To each of those sample distributions, we apply the evidence 
approximation method. We thus obtain 500 samples for the estimated Bayes factor, $B_e^s$, 
and for the estimated standard deviation, $\sigma_e^s$ (not including the correction factor). 

For each equation of state, we compute the distribution for 
$C_e = (B_e^s - B_e) / \bar{\sigma}_e$, 
i.e. the true error of the approximated Bayes factor normalized to the mean 
$\bar{\sigma}_e$ of the error estimate $\sigma_e^s$. 
We further combine the sample distributions $C_e$ for the different equations of state
into a single distribution $C$, using equal weights. This distribution is roughly
Gaussian and centered around zero. The one-sigma interval is $|C|<1.24$, and in 
89.3\% of the cases $|C|<2$. Hence, our chosen correction factor of 2 is larger 
than required for the example tested here. This is an added precaution owed to the 
{\em ad hoc} choice of the test case.

\bibliography{references}

\begin{thebibliography}{46}%
\makeatletter
\providecommand \@ifxundefined [1]{%
 \@ifx{#1\undefined}
}%
\providecommand \@ifnum [1]{%
 \ifnum #1\expandafter \@firstoftwo
 \else \expandafter \@secondoftwo
 \fi
}%
\providecommand \@ifx [1]{%
 \ifx #1\expandafter \@firstoftwo
 \else \expandafter \@secondoftwo
 \fi
}%
\providecommand \natexlab [1]{#1}%
\providecommand \enquote  [1]{``#1''}%
\providecommand \bibnamefont  [1]{#1}%
\providecommand \bibfnamefont [1]{#1}%
\providecommand \citenamefont [1]{#1}%
\providecommand \href@noop [0]{\@secondoftwo}%
\providecommand \href [0]{\begingroup \@sanitize@url \@href}%
\providecommand \@href[1]{\@@startlink{#1}\@@href}%
\providecommand \@@href[1]{\endgroup#1\@@endlink}%
\providecommand \@sanitize@url [0]{\catcode `\\12\catcode `\$12\catcode
  `\&12\catcode `\#12\catcode `\^12\catcode `\_12\catcode `\%12\relax}%
\providecommand \@@startlink[1]{}%
\providecommand \@@endlink[0]{}%
\providecommand \url  [0]{\begingroup\@sanitize@url \@url }%
\providecommand \@url [1]{\endgroup\@href {#1}{\urlprefix }}%
\providecommand \urlprefix  [0]{URL }%
\providecommand \Eprint [0]{\href }%
\providecommand \doibase [0]{http://dx.doi.org/}%
\providecommand \selectlanguage [0]{\@gobble}%
\providecommand \bibinfo  [0]{\@secondoftwo}%
\providecommand \bibfield  [0]{\@secondoftwo}%
\providecommand \translation [1]{[#1]}%
\providecommand \BibitemOpen [0]{}%
\providecommand \bibitemStop [0]{}%
\providecommand \bibitemNoStop [0]{.\EOS\space}%
\providecommand \EOS [0]{\spacefactor3000\relax}%
\providecommand \BibitemShut  [1]{\csname bibitem#1\endcsname}%
\let\auto@bib@innerbib\@empty
\bibitem [{\citenamefont {Oppenheimer}\ and\ \citenamefont
  {Volkoff}(1939)}]{PhysRev.55.374}%
  \BibitemOpen
  \bibfield  {author} {\bibinfo {author} {\bibfnamefont {J.~R.}\ \bibnamefont
  {Oppenheimer}}\ and\ \bibinfo {author} {\bibfnamefont {G.~M.}\ \bibnamefont
  {Volkoff}},\ }\href {\doibase 10.1103/PhysRev.55.374} {\bibfield  {journal}
  {\bibinfo  {journal} {Phys. Rev.}\ }\textbf {\bibinfo {volume} {55}},\
  \bibinfo {pages} {374} (\bibinfo {year} {1939})}\BibitemShut {NoStop}%
\bibitem [{\citenamefont {{\"O}zel}\ and\ \citenamefont
  {Freire}(2016)}]{annurev-astro-081915-023322}%
  \BibitemOpen
  \bibfield  {author} {\bibinfo {author} {\bibfnamefont {F.}~\bibnamefont
  {{\"O}zel}}\ and\ \bibinfo {author} {\bibfnamefont {P.}~\bibnamefont
  {Freire}},\ }\href {\doibase 10.1146/annurev-astro-081915-023322} {\bibfield
  {journal} {\bibinfo  {journal} {Annual Review of Astronomy and Astrophysics}\
  }\textbf {\bibinfo {volume} {54}},\ \bibinfo {pages} {401} (\bibinfo {year}
  {2016})},\ \Eprint
  {http://arxiv.org/abs/https://doi.org/10.1146/annurev-astro-081915-023322}
  {https://doi.org/10.1146/annurev-astro-081915-023322} \BibitemShut {NoStop}%
\bibitem [{\citenamefont {{Lindblom}}(1992)}]{1992ApJ...398..569L}%
  \BibitemOpen
  \bibfield  {author} {\bibinfo {author} {\bibfnamefont {L.}~\bibnamefont
  {{Lindblom}}},\ }\href {\doibase 10.1086/171882} {\bibfield  {journal}
  {\bibinfo  {journal} {The Astrophysical Journal}\ }\textbf {\bibinfo {volume}
  {398}},\ \bibinfo {pages} {569} (\bibinfo {year} {1992})}\BibitemShut
  {NoStop}%
\bibitem [{\citenamefont {Manchester}\ \emph {et~al.}(2005)\citenamefont
  {Manchester}, \citenamefont {Hobbs}, \citenamefont {Teoh},\ and\
  \citenamefont {Hobbs}}]{Manchester_2005}%
  \BibitemOpen
  \bibfield  {author} {\bibinfo {author} {\bibfnamefont {R.~N.}\ \bibnamefont
  {Manchester}}, \bibinfo {author} {\bibfnamefont {G.~B.}\ \bibnamefont
  {Hobbs}}, \bibinfo {author} {\bibfnamefont {A.}~\bibnamefont {Teoh}}, \ and\
  \bibinfo {author} {\bibfnamefont {M.}~\bibnamefont {Hobbs}},\ }\href
  {\doibase 10.1086/428488} {\bibfield  {journal} {\bibinfo  {journal} {The
  Astronomical Journal}\ }\textbf {\bibinfo {volume} {129}},\ \bibinfo {pages}
  {1993} (\bibinfo {year} {2005})}\BibitemShut {NoStop}%
\bibitem [{\citenamefont {ATNF-CSIRO}(2019)}]{NSCat}%
  \BibitemOpen
  \bibfield  {author} {\bibinfo {author} {\bibnamefont {ATNF-CSIRO}},\
  }\href@noop {} {\enquote {\bibinfo {title} {{ATNF Pulsar Catalog}},}\
  }\bibinfo {howpublished}
  {\url{https://www.atnf.csiro.au/research/pulsar/psrcat/}} (\bibinfo {year}
  {2019}),\ \bibinfo {note} {[Online; accessed 19-Nov-2019]}\BibitemShut
  {NoStop}%
\bibitem [{\citenamefont {NASA}(2019)}]{NICER}%
  \BibitemOpen
  \bibfield  {author} {\bibinfo {author} {\bibnamefont {NASA}},\ }\href@noop {}
  {\enquote {\bibinfo {title} {{NICER}},}\ }\bibinfo {howpublished}
  {\url{https://www.nasa.gov/nicer}} (\bibinfo {year} {2019}),\ \bibinfo {note}
  {[Online; accessed 26-Nov-2019]}\BibitemShut {NoStop}%
\bibitem [{\citenamefont {Miller}\ \emph {et~al.}(2019)\citenamefont {Miller}
  \emph {et~al.}}]{Miller:2019cac}%
  \BibitemOpen
  \bibfield  {author} {\bibinfo {author} {\bibfnamefont {M.~C.}\ \bibnamefont
  {Miller}} \emph {et~al.},\ }\href {\doibase 10.3847/2041-8213/ab50c5}
  {\bibfield  {journal} {\bibinfo  {journal} {Astrophys. J. Lett.}\ }\textbf
  {\bibinfo {volume} {887}},\ \bibinfo {pages} {L24} (\bibinfo {year}
  {2019})},\ \Eprint {http://arxiv.org/abs/1912.05705} {arXiv:1912.05705
  [astro-ph.HE]} \BibitemShut {NoStop}%
\bibitem [{\citenamefont {Riley}\ \emph {et~al.}(2019)\citenamefont {Riley},
  \citenamefont {Watts}, \citenamefont {Bogdanov}, \citenamefont {Ray},
  \citenamefont {Ludlam}, \citenamefont {Guillot}, \citenamefont {Arzoumanian},
  \citenamefont {Baker}, \citenamefont {Bilous}, \citenamefont {Chakrabarty},
  \citenamefont {Gendreau}, \citenamefont {Harding}, \citenamefont {Ho},
  \citenamefont {Lattimer}, \citenamefont {Morsink},\ and\ \citenamefont
  {Strohmayer}}]{Riley_2019}%
  \BibitemOpen
  \bibfield  {author} {\bibinfo {author} {\bibfnamefont {T.~E.}\ \bibnamefont
  {Riley}}, \bibinfo {author} {\bibfnamefont {A.~L.}\ \bibnamefont {Watts}},
  \bibinfo {author} {\bibfnamefont {S.}~\bibnamefont {Bogdanov}}, \bibinfo
  {author} {\bibfnamefont {P.~S.}\ \bibnamefont {Ray}}, \bibinfo {author}
  {\bibfnamefont {R.~M.}\ \bibnamefont {Ludlam}}, \bibinfo {author}
  {\bibfnamefont {S.}~\bibnamefont {Guillot}}, \bibinfo {author} {\bibfnamefont
  {Z.}~\bibnamefont {Arzoumanian}}, \bibinfo {author} {\bibfnamefont {C.~L.}\
  \bibnamefont {Baker}}, \bibinfo {author} {\bibfnamefont {A.~V.}\ \bibnamefont
  {Bilous}}, \bibinfo {author} {\bibfnamefont {D.}~\bibnamefont {Chakrabarty}},
  \bibinfo {author} {\bibfnamefont {K.~C.}\ \bibnamefont {Gendreau}}, \bibinfo
  {author} {\bibfnamefont {A.~K.}\ \bibnamefont {Harding}}, \bibinfo {author}
  {\bibfnamefont {W.~C.~G.}\ \bibnamefont {Ho}}, \bibinfo {author}
  {\bibfnamefont {J.~M.}\ \bibnamefont {Lattimer}}, \bibinfo {author}
  {\bibfnamefont {S.~M.}\ \bibnamefont {Morsink}}, \ and\ \bibinfo {author}
  {\bibfnamefont {T.~E.}\ \bibnamefont {Strohmayer}},\ }\href {\doibase
  10.3847/2041-8213/ab481c} {\bibfield  {journal} {\bibinfo  {journal} {The
  Astrophysical Journal}\ }\textbf {\bibinfo {volume} {887}},\ \bibinfo {pages}
  {L21} (\bibinfo {year} {2019})}\BibitemShut {NoStop}%
\bibitem [{\citenamefont {Aasi}\ \emph {et~al.}(2015)\citenamefont {Aasi} \emph
  {et~al.}}]{TheLIGOScientific:2014jea}%
  \BibitemOpen
  \bibfield  {author} {\bibinfo {author} {\bibfnamefont {J.}~\bibnamefont
  {Aasi}} \emph {et~al.} (\bibinfo {collaboration} {LIGO Scientific}),\ }\href
  {\doibase 10.1088/0264-9381/32/7/074001} {\bibfield  {journal} {\bibinfo
  {journal} {Class. Quant. Grav.}\ }\textbf {\bibinfo {volume} {32}},\ \bibinfo
  {pages} {074001} (\bibinfo {year} {2015})},\ \Eprint
  {http://arxiv.org/abs/1411.4547} {arXiv:1411.4547 [gr-qc]} \BibitemShut
  {NoStop}%
\bibitem [{\citenamefont {Acernese}\ \emph {et~al.}(2015)\citenamefont
  {Acernese} \emph {et~al.}}]{TheVirgo:2014hva}%
  \BibitemOpen
  \bibfield  {author} {\bibinfo {author} {\bibfnamefont {F.}~\bibnamefont
  {Acernese}} \emph {et~al.} (\bibinfo {collaboration} {VIRGO}),\ }\href
  {\doibase 10.1088/0264-9381/32/2/024001} {\bibfield  {journal} {\bibinfo
  {journal} {Class. Quant. Grav.}\ }\textbf {\bibinfo {volume} {32}},\ \bibinfo
  {pages} {024001} (\bibinfo {year} {2015})},\ \Eprint
  {http://arxiv.org/abs/1408.3978} {arXiv:1408.3978 [gr-qc]} \BibitemShut
  {NoStop}%
\bibitem [{\citenamefont {Poisson}(1998)}]{PhysRevD.57.5287}%
  \BibitemOpen
  \bibfield  {author} {\bibinfo {author} {\bibfnamefont {E.}~\bibnamefont
  {Poisson}},\ }\href {\doibase 10.1103/PhysRevD.57.5287} {\bibfield  {journal}
  {\bibinfo  {journal} {Phys. Rev. D}\ }\textbf {\bibinfo {volume} {57}},\
  \bibinfo {pages} {5287} (\bibinfo {year} {1998})}\BibitemShut {NoStop}%
\bibitem [{\citenamefont {Hinderer}(2008)}]{Hinderer_2008}%
  \BibitemOpen
  \bibfield  {author} {\bibinfo {author} {\bibfnamefont {T.}~\bibnamefont
  {Hinderer}},\ }\href {\doibase 10.1086/533487} {\bibfield  {journal}
  {\bibinfo  {journal} {The Astrophysical Journal}\ }\textbf {\bibinfo {volume}
  {677}},\ \bibinfo {pages} {1216} (\bibinfo {year} {2008})}\BibitemShut
  {NoStop}%
\bibitem [{\citenamefont {Flanagan}\ and\ \citenamefont
  {Hinderer}(2008)}]{Flanagan:2007ix}%
  \BibitemOpen
  \bibfield  {author} {\bibinfo {author} {\bibfnamefont {E.~E.}\ \bibnamefont
  {Flanagan}}\ and\ \bibinfo {author} {\bibfnamefont {T.}~\bibnamefont
  {Hinderer}},\ }\href {\doibase 10.1103/PhysRevD.77.021502} {\bibfield
  {journal} {\bibinfo  {journal} {Phys. Rev.}\ }\textbf {\bibinfo {volume}
  {D77}},\ \bibinfo {pages} {021502} (\bibinfo {year} {2008})},\ \Eprint
  {http://arxiv.org/abs/0709.1915} {arXiv:0709.1915 [astro-ph]} \BibitemShut
  {NoStop}%
\bibitem [{\citenamefont {Read}\ \emph
  {et~al.}(2009{\natexlab{a}})\citenamefont {Read}, \citenamefont {Markakis},
  \citenamefont {Shibata}, \citenamefont {Uryu}, \citenamefont {Creighton},\
  and\ \citenamefont {Friedman}}]{Read:2009yp}%
  \BibitemOpen
  \bibfield  {author} {\bibinfo {author} {\bibfnamefont {J.~S.}\ \bibnamefont
  {Read}}, \bibinfo {author} {\bibfnamefont {C.}~\bibnamefont {Markakis}},
  \bibinfo {author} {\bibfnamefont {M.}~\bibnamefont {Shibata}}, \bibinfo
  {author} {\bibfnamefont {K.}~\bibnamefont {Uryu}}, \bibinfo {author}
  {\bibfnamefont {J.~D.~E.}\ \bibnamefont {Creighton}}, \ and\ \bibinfo
  {author} {\bibfnamefont {J.~L.}\ \bibnamefont {Friedman}},\ }\href {\doibase
  10.1103/PhysRevD.79.124033} {\bibfield  {journal} {\bibinfo  {journal} {Phys.
  Rev.}\ }\textbf {\bibinfo {volume} {D79}},\ \bibinfo {pages} {124033}
  (\bibinfo {year} {2009}{\natexlab{a}})},\ \Eprint
  {http://arxiv.org/abs/0901.3258} {arXiv:0901.3258 [gr-qc]} \BibitemShut
  {NoStop}%
\bibitem [{\citenamefont {Del~Pozzo}\ \emph {et~al.}(2013)\citenamefont
  {Del~Pozzo}, \citenamefont {Li}, \citenamefont {Agathos}, \citenamefont {Van
  Den~Broeck},\ and\ \citenamefont {Vitale}}]{DelPozzo:2013ala}%
  \BibitemOpen
  \bibfield  {author} {\bibinfo {author} {\bibfnamefont {W.}~\bibnamefont
  {Del~Pozzo}}, \bibinfo {author} {\bibfnamefont {T.~G.~F.}\ \bibnamefont
  {Li}}, \bibinfo {author} {\bibfnamefont {M.}~\bibnamefont {Agathos}},
  \bibinfo {author} {\bibfnamefont {C.}~\bibnamefont {Van Den~Broeck}}, \ and\
  \bibinfo {author} {\bibfnamefont {S.}~\bibnamefont {Vitale}},\ }\href
  {\doibase 10.1103/PhysRevLett.111.071101} {\bibfield  {journal} {\bibinfo
  {journal} {Phys. Rev. Lett.}\ }\textbf {\bibinfo {volume} {111}},\ \bibinfo
  {pages} {071101} (\bibinfo {year} {2013})},\ \Eprint
  {http://arxiv.org/abs/1307.8338} {arXiv:1307.8338 [gr-qc]} \BibitemShut
  {NoStop}%
\bibitem [{\citenamefont {Agathos}\ \emph {et~al.}(2015)\citenamefont
  {Agathos}, \citenamefont {Meidam}, \citenamefont {Del~Pozzo}, \citenamefont
  {Li}, \citenamefont {Tompitak}, \citenamefont {Veitch}, \citenamefont
  {Vitale},\ and\ \citenamefont {Van Den~Broeck}}]{Agathos:2015uaa}%
  \BibitemOpen
  \bibfield  {author} {\bibinfo {author} {\bibfnamefont {M.}~\bibnamefont
  {Agathos}}, \bibinfo {author} {\bibfnamefont {J.}~\bibnamefont {Meidam}},
  \bibinfo {author} {\bibfnamefont {W.}~\bibnamefont {Del~Pozzo}}, \bibinfo
  {author} {\bibfnamefont {T.~G.~F.}\ \bibnamefont {Li}}, \bibinfo {author}
  {\bibfnamefont {M.}~\bibnamefont {Tompitak}}, \bibinfo {author}
  {\bibfnamefont {J.}~\bibnamefont {Veitch}}, \bibinfo {author} {\bibfnamefont
  {S.}~\bibnamefont {Vitale}}, \ and\ \bibinfo {author} {\bibfnamefont
  {C.}~\bibnamefont {Van Den~Broeck}},\ }\href {\doibase
  10.1103/PhysRevD.92.023012} {\bibfield  {journal} {\bibinfo  {journal} {Phys.
  Rev.}\ }\textbf {\bibinfo {volume} {D92}},\ \bibinfo {pages} {023012}
  (\bibinfo {year} {2015})},\ \Eprint {http://arxiv.org/abs/1503.05405}
  {arXiv:1503.05405 [gr-qc]} \BibitemShut {NoStop}%
\bibitem [{\citenamefont {van~der Sluys}\ \emph {et~al.}(2009)\citenamefont
  {van~der Sluys}, \citenamefont {Mandel}, \citenamefont {Raymond},
  \citenamefont {Kalogera}, \citenamefont {R{\"o}ver},\ and\ \citenamefont
  {Christensen}}]{van_der_Sluys_2009}%
  \BibitemOpen
  \bibfield  {author} {\bibinfo {author} {\bibfnamefont {M.}~\bibnamefont
  {van~der Sluys}}, \bibinfo {author} {\bibfnamefont {I.}~\bibnamefont
  {Mandel}}, \bibinfo {author} {\bibfnamefont {V.}~\bibnamefont {Raymond}},
  \bibinfo {author} {\bibfnamefont {V.}~\bibnamefont {Kalogera}}, \bibinfo
  {author} {\bibfnamefont {C.}~\bibnamefont {R{\"o}ver}}, \ and\ \bibinfo
  {author} {\bibfnamefont {N.}~\bibnamefont {Christensen}},\ }\href {\doibase
  10.1088/0264-9381/26/20/204010} {\bibfield  {journal} {\bibinfo  {journal}
  {Classical and Quantum Gravity}\ }\textbf {\bibinfo {volume} {26}},\ \bibinfo
  {pages} {204010} (\bibinfo {year} {2009})}\BibitemShut {NoStop}%
\bibitem [{\citenamefont {Veitch}\ and\ \citenamefont
  {Vecchio}(2010)}]{PhysRevD.81.062003}%
  \BibitemOpen
  \bibfield  {author} {\bibinfo {author} {\bibfnamefont {J.}~\bibnamefont
  {Veitch}}\ and\ \bibinfo {author} {\bibfnamefont {A.}~\bibnamefont
  {Vecchio}},\ }\href {\doibase 10.1103/PhysRevD.81.062003} {\bibfield
  {journal} {\bibinfo  {journal} {Phys. Rev. D}\ }\textbf {\bibinfo {volume}
  {81}},\ \bibinfo {pages} {062003} (\bibinfo {year} {2010})}\BibitemShut
  {NoStop}%
\bibitem [{\citenamefont {Raymond}\ \emph {et~al.}(2010)\citenamefont
  {Raymond}, \citenamefont {van~der Sluys}, \citenamefont {Mandel},
  \citenamefont {Kalogera}, \citenamefont {Röver},\ and\ \citenamefont
  {Christensen}}]{Raymond_2010}%
  \BibitemOpen
  \bibfield  {author} {\bibinfo {author} {\bibfnamefont {V.}~\bibnamefont
  {Raymond}}, \bibinfo {author} {\bibfnamefont {M.~V.}\ \bibnamefont {van~der
  Sluys}}, \bibinfo {author} {\bibfnamefont {I.}~\bibnamefont {Mandel}},
  \bibinfo {author} {\bibfnamefont {V.}~\bibnamefont {Kalogera}}, \bibinfo
  {author} {\bibfnamefont {C.}~\bibnamefont {Röver}}, \ and\ \bibinfo {author}
  {\bibfnamefont {N.}~\bibnamefont {Christensen}},\ }\href {\doibase
  10.1088/0264-9381/27/11/114009} {\bibfield  {journal} {\bibinfo  {journal}
  {Classical and Quantum Gravity}\ }\textbf {\bibinfo {volume} {27}},\ \bibinfo
  {pages} {114009} (\bibinfo {year} {2010})}\BibitemShut {NoStop}%
\bibitem [{\citenamefont {Rodriguez}\ \emph {et~al.}(2014)\citenamefont
  {Rodriguez}, \citenamefont {Farr}, \citenamefont {Raymond}, \citenamefont
  {Farr}, \citenamefont {Littenberg}, \citenamefont {Fazi},\ and\ \citenamefont
  {Kalogera}}]{Rodriguez_2014}%
  \BibitemOpen
  \bibfield  {author} {\bibinfo {author} {\bibfnamefont {C.~L.}\ \bibnamefont
  {Rodriguez}}, \bibinfo {author} {\bibfnamefont {B.}~\bibnamefont {Farr}},
  \bibinfo {author} {\bibfnamefont {V.}~\bibnamefont {Raymond}}, \bibinfo
  {author} {\bibfnamefont {W.~M.}\ \bibnamefont {Farr}}, \bibinfo {author}
  {\bibfnamefont {T.~B.}\ \bibnamefont {Littenberg}}, \bibinfo {author}
  {\bibfnamefont {D.}~\bibnamefont {Fazi}}, \ and\ \bibinfo {author}
  {\bibfnamefont {V.}~\bibnamefont {Kalogera}},\ }\href {\doibase
  10.1088/0004-637x/784/2/119} {\bibfield  {journal} {\bibinfo  {journal} {The
  Astrophysical Journal}\ }\textbf {\bibinfo {volume} {784}},\ \bibinfo {pages}
  {119} (\bibinfo {year} {2014})}\BibitemShut {NoStop}%
\bibitem [{\citenamefont {Veitch}\ \emph {et~al.}(2015)\citenamefont {Veitch}
  \emph {et~al.}}]{PhysRevD.91.042003}%
  \BibitemOpen
  \bibfield  {author} {\bibinfo {author} {\bibfnamefont {J.}~\bibnamefont
  {Veitch}} \emph {et~al.},\ }\href {\doibase 10.1103/PhysRevD.91.042003}
  {\bibfield  {journal} {\bibinfo  {journal} {Phys. Rev. D}\ }\textbf {\bibinfo
  {volume} {91}},\ \bibinfo {pages} {042003} (\bibinfo {year}
  {2015})}\BibitemShut {NoStop}%
\bibitem [{\citenamefont {Abbott}\ \emph {et~al.}(2019)\citenamefont {Abbott}
  \emph {et~al.}}]{PhysRevX.9.011001}%
  \BibitemOpen
  \bibfield  {author} {\bibinfo {author} {\bibfnamefont {B.~P.}\ \bibnamefont
  {Abbott}} \emph {et~al.} (\bibinfo {collaboration} {LIGO Scientific
  Collaboration and Virgo Collaboration}),\ }\href {\doibase
  10.1103/PhysRevX.9.011001} {\bibfield  {journal} {\bibinfo  {journal} {Phys.
  Rev. X}\ }\textbf {\bibinfo {volume} {9}},\ \bibinfo {pages} {011001}
  (\bibinfo {year} {2019})}\BibitemShut {NoStop}%
\bibitem [{\citenamefont {Abbott}\ \emph {et~al.}(2018)\citenamefont {Abbott}
  \emph {et~al.}}]{PhysRevLett.121.161101}%
  \BibitemOpen
  \bibfield  {author} {\bibinfo {author} {\bibfnamefont {B.~P.}\ \bibnamefont
  {Abbott}} \emph {et~al.} (\bibinfo {collaboration} {The LIGO Scientific
  Collaboration and the Virgo Collaboration}),\ }\href {\doibase
  10.1103/PhysRevLett.121.161101} {\bibfield  {journal} {\bibinfo  {journal}
  {Phys. Rev. Lett.}\ }\textbf {\bibinfo {volume} {121}},\ \bibinfo {pages}
  {161101} (\bibinfo {year} {2018})}\BibitemShut {NoStop}%
\bibitem [{\citenamefont {De}\ \emph {et~al.}(2018)\citenamefont {De},
  \citenamefont {Finstad}, \citenamefont {Lattimer}, \citenamefont {Brown},
  \citenamefont {Berger},\ and\ \citenamefont
  {Biwer}}]{PhysRevLett.121.091102}%
  \BibitemOpen
  \bibfield  {author} {\bibinfo {author} {\bibfnamefont {S.}~\bibnamefont
  {De}}, \bibinfo {author} {\bibfnamefont {D.}~\bibnamefont {Finstad}},
  \bibinfo {author} {\bibfnamefont {J.~M.}\ \bibnamefont {Lattimer}}, \bibinfo
  {author} {\bibfnamefont {D.~A.}\ \bibnamefont {Brown}}, \bibinfo {author}
  {\bibfnamefont {E.}~\bibnamefont {Berger}}, \ and\ \bibinfo {author}
  {\bibfnamefont {C.~M.}\ \bibnamefont {Biwer}},\ }\href {\doibase
  10.1103/PhysRevLett.121.091102} {\bibfield  {journal} {\bibinfo  {journal}
  {Phys. Rev. Lett.}\ }\textbf {\bibinfo {volume} {121}},\ \bibinfo {pages}
  {091102} (\bibinfo {year} {2018})}\BibitemShut {NoStop}%
\bibitem [{\citenamefont {Read}\ \emph
  {et~al.}(2009{\natexlab{b}})\citenamefont {Read}, \citenamefont {Lackey},
  \citenamefont {Owen},\ and\ \citenamefont {Friedman}}]{PhysRevD.79.124032}%
  \BibitemOpen
  \bibfield  {author} {\bibinfo {author} {\bibfnamefont {J.~S.}\ \bibnamefont
  {Read}}, \bibinfo {author} {\bibfnamefont {B.~D.}\ \bibnamefont {Lackey}},
  \bibinfo {author} {\bibfnamefont {B.~J.}\ \bibnamefont {Owen}}, \ and\
  \bibinfo {author} {\bibfnamefont {J.~L.}\ \bibnamefont {Friedman}},\ }\href
  {\doibase 10.1103/PhysRevD.79.124032} {\bibfield  {journal} {\bibinfo
  {journal} {Phys. Rev. D}\ }\textbf {\bibinfo {volume} {79}},\ \bibinfo
  {pages} {124032} (\bibinfo {year} {2009}{\natexlab{b}})}\BibitemShut
  {NoStop}%
\bibitem [{\citenamefont {Lindblom}(2010)}]{PhysRevD.82.103011}%
  \BibitemOpen
  \bibfield  {author} {\bibinfo {author} {\bibfnamefont {L.}~\bibnamefont
  {Lindblom}},\ }\href {\doibase 10.1103/PhysRevD.82.103011} {\bibfield
  {journal} {\bibinfo  {journal} {Phys. Rev. D}\ }\textbf {\bibinfo {volume}
  {82}},\ \bibinfo {pages} {103011} (\bibinfo {year} {2010})}\BibitemShut
  {NoStop}%
\bibitem [{\citenamefont {Landry}\ and\ \citenamefont
  {Essick}(2019)}]{Landry:2018prl}%
  \BibitemOpen
  \bibfield  {author} {\bibinfo {author} {\bibfnamefont {P.}~\bibnamefont
  {Landry}}\ and\ \bibinfo {author} {\bibfnamefont {R.}~\bibnamefont
  {Essick}},\ }\href {\doibase 10.1103/PhysRevD.99.084049} {\bibfield
  {journal} {\bibinfo  {journal} {Phys. Rev. D}\ }\textbf {\bibinfo {volume}
  {99}},\ \bibinfo {pages} {084049} (\bibinfo {year} {2019})},\ \Eprint
  {http://arxiv.org/abs/1811.12529} {arXiv:1811.12529 [gr-qc]} \BibitemShut
  {NoStop}%
\bibitem [{\citenamefont {Landry}\ \emph {et~al.}(2020)\citenamefont {Landry},
  \citenamefont {Essick},\ and\ \citenamefont
  {Chatziioannou}}]{Landry:2020vaw}%
  \BibitemOpen
  \bibfield  {author} {\bibinfo {author} {\bibfnamefont {P.}~\bibnamefont
  {Landry}}, \bibinfo {author} {\bibfnamefont {R.}~\bibnamefont {Essick}}, \
  and\ \bibinfo {author} {\bibfnamefont {K.}~\bibnamefont {Chatziioannou}},\
  }\href {\doibase 10.1103/PhysRevD.101.123007} {\bibfield  {journal} {\bibinfo
   {journal} {Phys. Rev. D}\ }\textbf {\bibinfo {volume} {101}},\ \bibinfo
  {pages} {123007} (\bibinfo {year} {2020})},\ \Eprint
  {http://arxiv.org/abs/2003.04880} {arXiv:2003.04880 [astro-ph.HE]}
  \BibitemShut {NoStop}%
\bibitem [{\citenamefont {Capano}\ \emph {et~al.}(2020)\citenamefont {Capano},
  \citenamefont {Tews}, \citenamefont {Brown}, \citenamefont {Margalit},
  \citenamefont {De}, \citenamefont {Kumar}, \citenamefont {Brown},
  \citenamefont {Krishnan},\ and\ \citenamefont {Reddy}}]{Capano:2019eae}%
  \BibitemOpen
  \bibfield  {author} {\bibinfo {author} {\bibfnamefont {C.~D.}\ \bibnamefont
  {Capano}}, \bibinfo {author} {\bibfnamefont {I.}~\bibnamefont {Tews}},
  \bibinfo {author} {\bibfnamefont {S.~M.}\ \bibnamefont {Brown}}, \bibinfo
  {author} {\bibfnamefont {B.}~\bibnamefont {Margalit}}, \bibinfo {author}
  {\bibfnamefont {S.}~\bibnamefont {De}}, \bibinfo {author} {\bibfnamefont
  {S.}~\bibnamefont {Kumar}}, \bibinfo {author} {\bibfnamefont {D.~A.}\
  \bibnamefont {Brown}}, \bibinfo {author} {\bibfnamefont {B.}~\bibnamefont
  {Krishnan}}, \ and\ \bibinfo {author} {\bibfnamefont {S.}~\bibnamefont
  {Reddy}},\ }\href {\doibase 10.1038/s41550-020-1014-6} {\bibfield  {journal}
  {\bibinfo  {journal} {Nature Astron.}\ }\textbf {\bibinfo {volume} {4}},\
  \bibinfo {pages} {625} (\bibinfo {year} {2020})},\ \Eprint
  {http://arxiv.org/abs/1908.10352} {arXiv:1908.10352 [astro-ph.HE]}
  \BibitemShut {NoStop}%
\bibitem [{\citenamefont {Dietrich}\ \emph {et~al.}(2020)\citenamefont
  {Dietrich}, \citenamefont {Coughlin}, \citenamefont {Pang}, \citenamefont
  {Bulla}, \citenamefont {Heinzel}, \citenamefont {Issa}, \citenamefont
  {Tews},\ and\ \citenamefont {Antier}}]{Dietrich1450}%
  \BibitemOpen
  \bibfield  {author} {\bibinfo {author} {\bibfnamefont {T.}~\bibnamefont
  {Dietrich}}, \bibinfo {author} {\bibfnamefont {M.~W.}\ \bibnamefont
  {Coughlin}}, \bibinfo {author} {\bibfnamefont {P.~T.~H.}\ \bibnamefont
  {Pang}}, \bibinfo {author} {\bibfnamefont {M.}~\bibnamefont {Bulla}},
  \bibinfo {author} {\bibfnamefont {J.}~\bibnamefont {Heinzel}}, \bibinfo
  {author} {\bibfnamefont {L.}~\bibnamefont {Issa}}, \bibinfo {author}
  {\bibfnamefont {I.}~\bibnamefont {Tews}}, \ and\ \bibinfo {author}
  {\bibfnamefont {S.}~\bibnamefont {Antier}},\ }\href {\doibase
  10.1126/science.abb4317} {\bibfield  {journal} {\bibinfo  {journal}
  {Science}\ }\textbf {\bibinfo {volume} {370}},\ \bibinfo {pages} {1450}
  (\bibinfo {year} {2020})},\ \Eprint
  {http://arxiv.org/abs/https://science.sciencemag.org/content/370/6523/1450.full.pdf}
  {https://science.sciencemag.org/content/370/6523/1450.full.pdf} \BibitemShut
  {NoStop}%
\bibitem [{\citenamefont {Abbott}\ \emph
  {et~al.}(2020{\natexlab{a}})\citenamefont {Abbott} \emph
  {et~al.}}]{LIGOScientific:2019eut}%
  \BibitemOpen
  \bibfield  {author} {\bibinfo {author} {\bibfnamefont {B.~P.}\ \bibnamefont
  {Abbott}} \emph {et~al.} (\bibinfo {collaboration} {LIGO Scientific,
  Virgo}),\ }\href {\doibase 10.1088/1361-6382/ab5f7c} {\bibfield  {journal}
  {\bibinfo  {journal} {Class. Quant. Grav.}\ }\textbf {\bibinfo {volume}
  {37}},\ \bibinfo {pages} {045006} (\bibinfo {year} {2020}{\natexlab{a}})},\
  \Eprint {http://arxiv.org/abs/1908.01012} {arXiv:1908.01012 [gr-qc]}
  \BibitemShut {NoStop}%
\bibitem [{\citenamefont {Abbott}\ \emph {et~al.}(2021)\citenamefont {Abbott}
  \emph {et~al.}}]{RICHABBOTT2021100658}%
  \BibitemOpen
  \bibfield  {author} {\bibinfo {author} {\bibfnamefont {B.~P.}\ \bibnamefont
  {Abbott}} \emph {et~al.},\ }\href {\doibase
  https://doi.org/10.1016/j.softx.2021.100658} {\bibfield  {journal} {\bibinfo
  {journal} {SoftwareX}\ }\textbf {\bibinfo {volume} {13}},\ \bibinfo {pages}
  {100658} (\bibinfo {year} {2021})}\BibitemShut {NoStop}%
\bibitem [{\citenamefont {\mbox{LIGO Scientific Collaboration, Virgo
  Collaboration}}(2017)}]{gw170817}%
  \BibitemOpen
  \bibfield  {author} {\bibinfo {author} {\bibnamefont {\mbox{LIGO Scientific
  Collaboration, Virgo Collaboration}}},\ }\href {\doibase 10.7935/82H3-HH23}
  {\enquote {\bibinfo {title} {\mbox{GW170817}},}\ }\bibinfo {howpublished}
  {\url{https://www.gw-openscience.org/eventapi/html/GWTC-1-confident/GW170817/v3}}
  (\bibinfo {year} {2017})\BibitemShut {NoStop}%
\bibitem [{\citenamefont {{Skilling}}(2004)}]{2004AIPC..735..395S}%
  \BibitemOpen
  \bibfield  {author} {\bibinfo {author} {\bibfnamefont {J.}~\bibnamefont
  {{Skilling}}},\ }in\ \href {\doibase 10.1063/1.1835238} {\emph {\bibinfo
  {booktitle} {Bayesian Inference and Maximum Entropy Methods in Science and
  Engineering: 24th International Workshop on Bayesian Inference and Maximum
  Entropy Methods in Science and Engineering}}},\ \bibinfo {series} {American
  Institute of Physics Conference Series}, Vol.\ \bibinfo {volume} {735},\
  \bibinfo {editor} {edited by\ \bibinfo {editor} {\bibfnamefont
  {R.}~\bibnamefont {{Fischer}}}, \bibinfo {editor} {\bibfnamefont
  {R.}~\bibnamefont {{Preuss}}}, \ and\ \bibinfo {editor} {\bibfnamefont
  {U.~V.}\ \bibnamefont {{Toussaint}}}}\ (\bibinfo {year} {2004})\ pp.\
  \bibinfo {pages} {395--405}\BibitemShut {NoStop}%
\bibitem [{\citenamefont {Favata}(2014)}]{Favata:2013rwa}%
  \BibitemOpen
  \bibfield  {author} {\bibinfo {author} {\bibfnamefont {M.}~\bibnamefont
  {Favata}},\ }\href {\doibase 10.1103/PhysRevLett.112.101101} {\bibfield
  {journal} {\bibinfo  {journal} {Phys. Rev. Lett.}\ }\textbf {\bibinfo
  {volume} {112}},\ \bibinfo {pages} {101101} (\bibinfo {year} {2014})},\
  \Eprint {http://arxiv.org/abs/1310.8288} {arXiv:1310.8288 [gr-qc]}
  \BibitemShut {NoStop}%
\bibitem [{\citenamefont {Wade}\ \emph {et~al.}(2014)\citenamefont {Wade},
  \citenamefont {Creighton}, \citenamefont {Ochsner}, \citenamefont {Lackey},
  \citenamefont {Farr}, \citenamefont {Littenberg},\ and\ \citenamefont
  {Raymond}}]{Wade:2014vqa}%
  \BibitemOpen
  \bibfield  {author} {\bibinfo {author} {\bibfnamefont {L.}~\bibnamefont
  {Wade}}, \bibinfo {author} {\bibfnamefont {J.~D.~E.}\ \bibnamefont
  {Creighton}}, \bibinfo {author} {\bibfnamefont {E.}~\bibnamefont {Ochsner}},
  \bibinfo {author} {\bibfnamefont {B.~D.}\ \bibnamefont {Lackey}}, \bibinfo
  {author} {\bibfnamefont {B.~F.}\ \bibnamefont {Farr}}, \bibinfo {author}
  {\bibfnamefont {T.~B.}\ \bibnamefont {Littenberg}}, \ and\ \bibinfo {author}
  {\bibfnamefont {V.}~\bibnamefont {Raymond}},\ }\href {\doibase
  10.1103/PhysRevD.89.103012} {\bibfield  {journal} {\bibinfo  {journal} {Phys.
  Rev.}\ }\textbf {\bibinfo {volume} {D89}},\ \bibinfo {pages} {103012}
  (\bibinfo {year} {2014})},\ \Eprint {http://arxiv.org/abs/1402.5156}
  {arXiv:1402.5156 [gr-qc]} \BibitemShut {NoStop}%
\bibitem [{\citenamefont {Bini}\ \emph {et~al.}(2012)\citenamefont {Bini},
  \citenamefont {Damour},\ and\ \citenamefont {Faye}}]{Bini:2012gu}%
  \BibitemOpen
  \bibfield  {author} {\bibinfo {author} {\bibfnamefont {D.}~\bibnamefont
  {Bini}}, \bibinfo {author} {\bibfnamefont {T.}~\bibnamefont {Damour}}, \ and\
  \bibinfo {author} {\bibfnamefont {G.}~\bibnamefont {Faye}},\ }\href {\doibase
  10.1103/PhysRevD.85.124034} {\bibfield  {journal} {\bibinfo  {journal} {Phys.
  Rev. D}\ }\textbf {\bibinfo {volume} {85}},\ \bibinfo {pages} {124034}
  (\bibinfo {year} {2012})},\ \Eprint {http://arxiv.org/abs/1202.3565}
  {arXiv:1202.3565 [gr-qc]} \BibitemShut {NoStop}%
\bibitem [{\citenamefont {\mbox{LIGO Scientific Collaboration, Virgo
  Collaboration}}(2018)}]{lalsuite}%
  \BibitemOpen
  \bibfield  {author} {\bibinfo {author} {\bibnamefont {\mbox{LIGO Scientific
  Collaboration, Virgo Collaboration}}},\ }\href {\doibase 10.7935/GT1W-FZ16}
  {\enquote {\bibinfo {title} {\mbox{LALSuite}},}\ }\bibinfo {howpublished}
  {\url{https://git.ligo.org/lscsoft/lalsuite}} (\bibinfo {year}
  {2018})\BibitemShut {NoStop}%
\bibitem [{\citenamefont {Metropolis}\ \emph {et~al.}(1953)\citenamefont
  {Metropolis} \emph {et~al.}}]{Metrolis}%
  \BibitemOpen
  \bibfield  {author} {\bibinfo {author} {\bibfnamefont {N.}~\bibnamefont
  {Metropolis}} \emph {et~al.},\ }\href@noop {} {\bibfield  {journal} {\bibinfo
   {journal} {J. Chem. Phys.}\ }\textbf {\bibinfo {volume} {21}},\ \bibinfo
  {pages} {1087} (\bibinfo {year} {1953})}\BibitemShut {NoStop}%
\bibitem [{\citenamefont {Hastings}(1970)}]{Hastings}%
  \BibitemOpen
  \bibfield  {author} {\bibinfo {author} {\bibfnamefont {W.}~\bibnamefont
  {Hastings}},\ }\href@noop {} {\bibfield  {journal} {\bibinfo  {journal}
  {Biometrika}\ }\textbf {\bibinfo {volume} {57}},\ \bibinfo {pages} {97}
  (\bibinfo {year} {1970})}\BibitemShut {NoStop}%
\bibitem [{\citenamefont {Abbott}\ \emph
  {et~al.}(2020{\natexlab{b}})\citenamefont {Abbott} \emph
  {et~al.}}]{Abbott:2020uma}%
  \BibitemOpen
  \bibfield  {author} {\bibinfo {author} {\bibfnamefont {B.~P.}\ \bibnamefont
  {Abbott}} \emph {et~al.} (\bibinfo {collaboration} {LIGO Scientific,
  Virgo}),\ }\href {\doibase 10.3847/2041-8213/ab75f5} {\bibfield  {journal}
  {\bibinfo  {journal} {Astrophys. J. Lett.}\ }\textbf {\bibinfo {volume}
  {892}},\ \bibinfo {pages} {L3} (\bibinfo {year} {2020}{\natexlab{b}})},\
  \Eprint {http://arxiv.org/abs/2001.01761} {arXiv:2001.01761 [astro-ph.HE]}
  \BibitemShut {NoStop}%
\bibitem [{\citenamefont {\mbox{LIGO Scientific Collaboration, Virgo
  Collaboration}}(2019)}]{gw190425}%
  \BibitemOpen
  \bibfield  {author} {\bibinfo {author} {\bibnamefont {\mbox{LIGO Scientific
  Collaboration, Virgo Collaboration}}},\ }\href {\doibase 10.7935/ggb8-1v94}
  {\enquote {\bibinfo {title} {\mbox{GW190425}},}\ }\bibinfo {howpublished}
  {\url{https://www.gw-openscience.org/eventapi/html/O3_Discovery_Papers/GW190425/v1}}
  (\bibinfo {year} {2019})\BibitemShut {NoStop}%
\bibitem [{\citenamefont {Ghosh}(2020{\natexlab{a}})}]{gwxtreme-pypi}%
  \BibitemOpen
  \bibfield  {author} {\bibinfo {author} {\bibfnamefont {S.}~\bibnamefont
  {Ghosh}},\ }\href@noop {} {\enquote {\bibinfo {title} {{GWX}treme package},}\
  }\bibinfo {howpublished} {\url{https://pypi.org/project/GWXtreme/}} (\bibinfo
  {year} {2020}{\natexlab{a}}),\ \bibinfo {note} {[Online; accessed
  07-Jun-2020]}\BibitemShut {NoStop}%
\bibitem [{\citenamefont {Ghosh}(2020{\natexlab{b}})}]{gwxtreme-doc}%
  \BibitemOpen
  \bibfield  {author} {\bibinfo {author} {\bibfnamefont {S.}~\bibnamefont
  {Ghosh}},\ }\href@noop {} {\enquote {\bibinfo {title} {{GWX}treme
  documentation},}\ }\bibinfo {howpublished}
  {\url{https://gwxtreme.readthedocs.io/en/latest/}} (\bibinfo {year}
  {2020}{\natexlab{b}}),\ \bibinfo {note} {[Online; accessed
  07-Jun-2020]}\BibitemShut {NoStop}%
\bibitem [{\citenamefont {{Lackey}}\ and\ \citenamefont
  {{Wade}}(2015)}]{2015PhRvD..91d3002L}%
  \BibitemOpen
  \bibfield  {author} {\bibinfo {author} {\bibfnamefont {B.~D.}\ \bibnamefont
  {{Lackey}}}\ and\ \bibinfo {author} {\bibfnamefont {L.}~\bibnamefont
  {{Wade}}},\ }\href {\doibase 10.1103/PhysRevD.91.043002} {\bibfield
  {journal} {\bibinfo  {journal} {\prd}\ }\textbf {\bibinfo {volume} {91}},\
  \bibinfo {eid} {043002} (\bibinfo {year} {2015})},\ \Eprint
  {http://arxiv.org/abs/1410.8866} {arXiv:1410.8866 [gr-qc]} \BibitemShut
  {NoStop}%
\bibitem [{\citenamefont {Ghosh}\ \emph {et~al.}(2021)\citenamefont {Ghosh},
  \citenamefont {Liu}, \citenamefont {Creighton}, \citenamefont {Kastaun},
  \citenamefont {Pratten},\ and\ \citenamefont {Maga\~na}}]{dataset}%
  \BibitemOpen
  \bibfield  {author} {\bibinfo {author} {\bibfnamefont {S.}~\bibnamefont
  {Ghosh}}, \bibinfo {author} {\bibfnamefont {X.}~\bibnamefont {Liu}}, \bibinfo
  {author} {\bibfnamefont {J.}~\bibnamefont {Creighton}}, \bibinfo {author}
  {\bibfnamefont {W.}~\bibnamefont {Kastaun}}, \bibinfo {author} {\bibfnamefont
  {G.}~\bibnamefont {Pratten}}, \ and\ \bibinfo {author} {\bibfnamefont
  {I.}~\bibnamefont {Maga\~na}},\ }\href {\doibase 10.5281/zenodo.4679013}
  {\enquote {\bibinfo {title} {Dataset for rapid model comparison of equations
  of state from gravitational wave observation of binary neutron star
  coalescences},}\ } (\bibinfo {year} {2021})\BibitemShut {NoStop}%
\end{thebibliography}%

\end{document}